\documentclass[aps,prd,onecolumn,amsmath,longbibliography,amssymb,superscriptaddress]{revtex4}
\usepackage{graphicx}
\usepackage{subfigure}
\usepackage{epsfig}
\usepackage{bm}
\usepackage{ulem}
\usepackage{color}
\usepackage{multirow}
\usepackage{slashed}
\usepackage{hyperref}
\usepackage{verbatim}
\usepackage{braket}
\usepackage{url}
\usepackage{natbib}

\usepackage{tikz}
\usetikzlibrary{arrows,shapes}
\usetikzlibrary{trees}
\usetikzlibrary{matrix,arrows} 				
\usetikzlibrary{positioning}				
\usetikzlibrary{calc,through}				
\usetikzlibrary{decorations.pathreplacing}  
\usepackage{pgffor}							

\usetikzlibrary{decorations.pathmorphing}	
\usetikzlibrary{decorations.markings}
\tikzset{
	vector/.style={decorate, decoration={snake}, draw},
	provector/.style={decorate, decoration={snake,amplitude=2.5pt}, draw},
	antivector/.style={decorate, decoration={snake,amplitude=-2.5pt}, draw},
	fermion/.style={draw=black, postaction={decorate},
		decoration={markings,mark=at position .55 with {\arrow[draw=black]{>}}}},
	fermiona/.style={draw=red},
	fermionbar/.style={draw=black, postaction={decorate},
		decoration={markings,mark=at position .55 with {\arrow[draw=black]{<}}}},
	fermionnoarrow/.style={draw=black},
	gluon/.style={decorate, draw=black,
		decoration={coil,amplitude=4pt, segment length=5pt}},
	scalar/.style={dashed,draw=black, postaction={decorate},
		decoration={markings,mark=at position .55 with {\arrow[draw=black]{>}}}},
	scalarbar/.style={dashed,draw=black, postaction={decorate},
		decoration={markings,mark=at position .55 with {\arrow[draw=black]{<}}}},
	scalarnoarrow/.style={dashed,draw=black},
	electron/.style={draw=black, postaction={decorate},
		decoration={markings,mark=at position .55 with {\arrow[draw=black]{>}}}},
	bigvector/.style={decorate, decoration={snake,amplitude=4pt}, draw},
}

\hypersetup{colorlinks=true,linkcolor=blue,citecolor=blue,urlcolor=blue}

\def\ben{\begin{equation}}
	\def\een{\end{equation}}

\def\be{\begin{equation}}
	\def\ee{\end{equation}}
\def\beq{\begin{equation}}
	\def\eeq{\end{equation}}
\def\ba{\begin{array}}
	\def\ea{\end{array}}

\def\dalemb#1#2{{\vbox{\hrule height .#2pt
			\hbox{\vrule width.#2pt height#1pt \kern#1pt
				\vrule width.#2pt}
			\hrule height.#2pt}}}

\newcommand{\bea}{\begin{eqnarray}}
	\newcommand{\eea}{\end{eqnarray}}

\DeclareDocumentCommand{\nint}{ O{} O{} m }{\ensuremath{ \int_{\mbox{\scriptsize $#1$}}^{\mbox{\scriptsize$#2$}}\!\!\! \mbox{\small $\,\mathrm{d}#3$\! }}}

\begin{document}
\title{Thermodynamic and Real-time Dynamic Properties of Complex Sachdev-Ye-Kitaev Model}
\author{Sizheng Cao}
\affiliation{Department of Physics, Shanghai University, Shangda Road 99, Baoshan, Shanghai, 200444, China}
\author{Xian-Hui Ge~\footnote{ Corresponding author. gexh@shu.edu.cn}}
\affiliation{Department of Physics, Shanghai University, Shangda Road 99, Baoshan, Shanghai, 200444, China}
\author{Yi-Cheng Rui}
\affiliation{Department of Physics, Shanghai University, Shangda Road 99, Baoshan, Shanghai, 200444, China}
\affiliation{Tsung-Dao Lee Institute, Shanghai Jiao Tong University, Shengrong Road 520, Shanghai, 201210, People’s Republic of China}
\affiliation{School of Physics and Astronomy, Shanghai Jiao Tong University, 800 Dongchuan Road, Shanghai 200240, People’s Republic of China}


\begin{abstract}
We study the complex Sachdev-Ye-Kitaev (cSYK)  numerically and investigate thermodynamic behavior of cSYK model across varying chemical potentials. We discover that the cSYK model remarkably mirrors the first-order phase transition seen in the van der Waals-Maxwell system, culminating at a non-mean-field critical point with distinctively different critical exponents. We analyze in detail the similarity between the van der Waals phase transition and the cSYK model, and further explore the mechanism by which the chemical potential drives the phase transition in the system. Exact diagonalization for the cSYK model reveals the significant impact of chemical potential on energy distribution, with observable energy gaps in the gapped phase. Quantum chaos indicators, including spectral form factors, suggest more stable energy states in the neutral case. Real-time dynamics, analyzed via analytical continuation of Schwinger-Dyson equations, show rapid decay in the gapless phase and prolonged oscillation lifetimes in the gapped regime. Spectral functions imply a shift from a continuous to a discrete energy level distribution, emphasizing the critical role of chemical potential  in shaping spectral properties.

\end{abstract}
\date{}
\maketitle
\section{INTRODUCTION}
The Sachdev-Ye-Kitaev (SYK) models, initially introduced in the context of quantum many-body systems \cite{SY93}, have garnered significant interest due to their rich phenomenology and profound connections to holography and quantum gravity \cite{kitaev15,stanford,Jensen2016,MQ18,garcia17,zhangpf,aze17,ysyk1,ysyk21,2006sahoo,1910.14099,Jian2021,2208-10800,cao2024excitation,PhysRevResearch.2.023366,Cai2018,garcíagarcía2023classification,garcíagarcía2024lyapunov}. These models are characterized by a set of fermions interacting through random couplings, which leads to emergent conformal symmetry and universal thermalization properties \cite{Dai_2019,Nayak2019}. Their simplicity in analytical treatment and their ability to capture non-trivial aspects of strongly correlated quantum systems make them ideal theoretical laboratories for exploring the interplay between quantum chaos, holography, and gravity.

The SYK models have been shown to possess a dual description in terms of two-dimensional anti-de Sitter space ($AdS_2$) \cite{Jensen2016}, which connects the quantum dynamics of the SYK model to classical gravitational systems. This correspondence, often referred to as the SYK/AdS duality \cite{wbfu,cp,bul,syk19,papadi,newboundary,godet,pankaj,Jan_PRD}, provides a concrete realization of the AdS/CFT correspondence in a simplified setting, allowing for a detailed investigation of the holographic principle at a microscopic level.

In this study, we focus on the complex Sachdev-Ye-Kitaev (cSYK) model, an extension of the original SYK model that incorporates complex fermions \cite{Cotler_2017,Fu_2016}. A complex fermion can be viewed as consisting of two real fermions, often described as its real and imaginary parts, which independently obey the Fermi-Dirac statistics and anticommute with each other. The cSYK model introduces additional degrees of freedom and interactions, leading to richer phase diagrams and more intricate thermodynamic behavior compared to the real SYK model. Our primary goal is to explore the thermodynamic and real-time dynamics of the cSYK model, specifically focusing on the phase transition temperatures, free energy, critical exponents, exact diagonalization, spectral form factors, real-time dynamics, and spectral functions.

The cSYK model exhibits a marked first-order phase transition from a gapped to a gapless phase as the chemical potential varies \cite{Cotler_2017}. This transition bears striking similarities to the phase transitions observed in the van der Waals-Maxwell system, culminating at a non-mean-field critical point with distinct critical exponents. The comparison with the van der Waals-Maxwell system is particularly intriguing since both systems display a first-order phase transition, indicating a potential universality class that extends beyond the specific details of each model.

To elucidate the phase structure of the cSYK model, we employ exact diagonalization techniques \cite{PhysRevB.97.235124,Cotler_2017,Fu_2016,Cheng2014,wei15} for finite $N$, where 
$N$ denotes the number of complex fermions. This approach allows us to access the full spectrum of the Hamiltonian and analyze the spectral density as a function of the chemical potential. Recognizing that the breaking of particle-hole symmetry induced by the chemical potential is the direct cause of the observed phase transition, we concretely study how chemical potential alters the properties of the model from the perspective of energy level distribution. We precisely study the impact of the chemical potential on the Hamiltonian's spectral density distribution. We observe that the chemical potential significantly influences the energy distribution, causing the shift and tilt of the spectral peak position, which breaks the original even symmetric spectral distribution $\rho(-E)=\rho(E)$.

From the thermodynamic study, we observed a clear first-order phase transition between the gapped and gapless phases driven by the chemical potential. To gain deeper insights into the role of the chemical potential in driving this transition, we extend the investigation into the real-time dynamics and spectral properties of the system. The spectral form factor (SFF) \cite{saad2019semiclassical,ALTLAND201845,PhysRevD.98.086026,PhysRevA.103.013323}, a key indicator of quantum chaos, is calculated to probe the stability of energy states in the cSYK model. The SFF reveals that energy states are more stable in the absence of additional particle or hole excitations, pointing towards a higher degree of coherence in the gapped phase compared to the gapless phase. Real-time dynamics \cite{2003-03914,2009-10759,Maldacena2021} and spectral functions are explored through the Schwinger-Dyson equations, offering insights into the response of the cSYK model to external perturbations. The greater Green's function exhibits a rapid decay in the gapless phase, indicative of short-lived excitations. Conversely, in the gapped phase, the oscillation lifetime is significantly extended, reflecting the presence of well-defined energy levels. Spectral function analysis confirms a continuous and broad distribution in the gapless phase, transitioning to a single peak in the gapped phase, highlighting the role of the chemical potential in driving the system from a continuous spectrum to discrete energy levels.

The remainder of this paper is structured as follows. Section \ref{sec:model} introduces the complex Sachdev-Ye-Kitaev model and outlines the methods employed for our numerical analysis, and analyzes the impact of particle-hole symmetry breaking on the model and the mechanism behind the phase transition. 
This section also presents the results of our investigation, including the phase diagram, free energy and critical exponents, where we describe in detail how the phase transition behavior of the cSYK model is analogous to a van der Waals transition. In Section \ref{ED}, we discuss the exact diagonalization and spectral form factor of the cSYK model. Section \ref{realtime} delves into the real-time dynamics and spectral functions of this system. The results reveal that in the low temperature limit, there is a transition
of the system from a continuous spectrum to discrete energy levels. The last section concludes the paper with a summary of our main findings and directions for future research. In appendix, we show the effect of chemical potential of $q=2$ complex SYK model.

\section{THE COMPLEX SYK MODEL AND THERMODYNAMICS}\label{sec:model}
We begin by considering the cSYK model, a variant of the Majorana SYK model that conserves charge. The Hamiltonian for this model is given by
\bea
H = \sum_{1 \leq i_1 \leq \cdots \leq i_q \leq N, 1 \leq j_1 \leq \cdots \leq j_q \leq N} J^{i_1 \cdots i_q}_{j_1 \cdots j_q} c^{\dagger}_{i_1} \cdots c^{\dagger}_{i_q} c_{j_1} \cdots c_{j_q} - \mu \sum_{i=1}^{N} c^{\dagger}_{i} c_{i},
\eea
where \( J^{i_1 \cdots i_q}_{j_1 \cdots j_q} \) are coupling constants and \( c_i, c^{\dagger}_i \) are fermion creation and annihilation operators.

In this study, we primarily examine the case where \( q = 4 \). The corresponding Hamiltonian simplifies to
\bea\label{H}
H = \sum_{i,j,k,l=1}^{N} J_{ijkl} c^{\dagger}_{i} c^{\dagger}_{j} c_k c_l - \mu \sum_{i=1}^{N} c^{\dagger}_{i} c_{i},
\eea
where \( c_i \) and \( c^{\dagger}_i \) are fermion operators that obey the anticommutation relation \( \{c_i, c^{\dagger}_j\} = \delta_{ij} \). The coupling constants \( J_{ijkl} \) are independent random complex Gaussian variables with zero mean and a variance given by
\bea
\overline{J_{ijkl}} = 0, \quad \sigma^2 = \overline{|J_{ijkl}|^2} = \frac{J^2}{8 N^3}.
\eea

We introduce the global conserved \( U(1) \) charge, defined as
\bea
Q = \sum_{i=1}^{N} \left(c^{\dagger}_i c_i - \frac{1}{2}\right).
\eea
This charge is linked to the ultraviolet (UV) asymmetry of the Green's function, defined as
\bea
G(\tau, \tau') = \langle T c_i(\tau) c^{\dagger}_i(\tau') \rangle, \quad G(0^-) = -\frac{1}{2} - \mathcal{Q}, \quad G(0^+) = \frac{1}{2} - \mathcal{Q},
\eea
where \( \mathcal{Q} = \frac{\langle Q \rangle}{N} \) the charge density of this system.

In the infrared (IR) limit, the Green's function is characterized by its long-time behavior
\bea
G_{\beta = \infty}(\pm \tau) \sim \mp e^{\pm \pi \mathcal{E}} \tau^{-1/2}, \quad \tau \gg J^{-1},
\eea
where \( \mathcal{E} \) is a parameter that governs the particle-hole symmetry.

\subsection{Replica trick and partition function}
The partition function is a critical tool for understanding the thermodynamic properties of a model. In the context of Euclidean time, the partition function for a system can be expressed as
\bea
Z = \mathrm{Tr}\left[e^{-\beta H}\right] = \int \prod_{i} \mathcal{D}c_i^\dagger \mathcal{D}c_i \ \exp \left\{ -\int_0^\beta d\tau \left(\sum_i c_i^\dagger \partial_\tau c_i + H\right) \right\}.
\eea
When we incorporate the Hamiltonian (\ref{H}) into this expression, we obtain
\bea
Z = \int \prod_{i} \mathcal{D}c_i^\dagger \mathcal{D}c_i \ \exp \left\{ -\int_0^\beta d\tau \left( \sum_{i=1}^N c_i^\dagger \partial_\tau c_i + \sum_{i,j,k,l=1}^{N} J_{ijkl} c^{\dagger}_{i} c^{\dagger}_{j} c_k c_l - \mu \sum_{i=1}^{N} c^{\dagger}_{i} c_{i} \right) \right\}.
\eea
Due to the involvement of random variables in the system, it is essential to focus on the average behavior. The average of an arbitrary function \( f(J_{ijkl}) \) of the random variables \( J_{ijkl} \) is defined as
\bea
\langle f(J_{ijkl}) \rangle_J = \int \prod_{i,j,k,l} dJ_{ijkl} \ P(J_{ijkl}) f(J_{ijkl}),
\eea
where the probability distribution \( P(J_{ijkl}) \) is given by
\bea
P(J_{ijkl}) = e^{-\frac{|J_{ijkl}|^2}{\sigma^2}} = e^{-\frac{|J_{ijkl}|^2 8N^3}{J^2}}.
\eea
where the subscript $J$ denotes the average over random variables $\sigma^2=J^2/(8N^3)$. This formulation allows us to statistically average over the random couplings \( J_{ijkl} \), thereby focusing on the typical behavior of the system rather than specific realizations.

Therefore, the average partition function can be formulated as
\bea
\langle Z \rangle_J = \int \prod_{i} \mathcal{D}c_i^\dagger \mathcal{D}c_i \ \prod_{i,j,k,l} dJ_{ijkl} P(J_{ijkl}) \exp \left\{ -\int_0^\beta d\tau \left( \sum_{i=1}^N c_i^\dagger \partial_\tau c_i + \sum_{i,j,k,l=1}^N J_{ijkl} c^{\dagger}_{i} c^{\dagger}_{j} c_k c_l - \mu \sum_{i=1}^N c^{\dagger}_{i} c_{i} \right) \right\}.
\eea
To determine the free energy, we use the relationship \( F = -T \ln Z \). To facilitate this calculation, we employ the replica trick, which rewrites the integral into a more tractable form. Specifically, this method involves
\bea
\langle F \rangle_J = -T \langle \ln Z \rangle_J = -T \lim_{M \rightarrow 0} \frac{\ln \langle Z^M \rangle_J}{M}.
\eea
By introducing the replica trick, we can simplify the process of averaging the logarithm of the partition function, ultimately allowing us to compute the system's average free energy.

In a physical context, the replica trick is a method where we duplicate the system into \( M \) identical replicas. Initially, we obtain an expression that depends on the integer \( M \). By employing analytical continuation, this expression is extended to arbitrary \( M \), and finally, we take the limit as \( M \) approaches zero. Although this approach might seem counterintuitive, it resolves the issue effectively. The integral \( \int P(J_{ijkl}) Z^M dJ_{ijkl} \) is significantly simpler to manage. A comprehensive example of applying the replica trick to the Majorana SYK model is provided in \cite{softmode}. The same method is applicable to the complex SYK model, which we will outline briefly.

We use the index \( m \) to label each of the \( M \) copies. The replicated partition function is then formulated as
\bea
\langle Z^M \rangle_J &=& \int \prod_{i} \mathcal{D}c_i^\dagger \mathcal{D}c_i \ \prod_{i,j,k,l} dJ_{ijkl} dJ_{ijkl}^* P(J_{ijkl})\exp[ -\sum_{m=1}^M \int_0^\beta d\tau \left( \sum_i c_{im}^\dagger (\partial_\tau - \mu) c_{im} + \sum_{i,j,k,l=1}^{N} J_{ijkl} c^{\dagger}_{im} c^{\dagger}_{jm} c_{km} c_{lm} \right)].\nonumber\\
\eea
This approach simplifies the averaging process by transforming the problem into one that involves integrals over the replicated variables, which are more manageable than dealing directly with the logarithm of the partition function.

The complex Gaussian integral differs somewhat from its real counterpart. The parts of the integral that depend on \( J_{ijkl} \) are given by
\bea\label{gaussian integral}
\int \prod_{ijkl} dJ_{ijkl} dJ_{ijkl}^* \exp \left[ -\frac{|J_{ijkl}|^2}{\sigma^2} \right] \exp \left[ -\sum_{ijkl} J_{ijkl} \int d\tau \ c_{im}^\dagger c_{jm}^\dagger c_{km} c_{lm} \right].
\eea
By following the method outlined in \cite{2006sahoo}, this integral can be transformed into
\bea
&& \int \mathcal{D}J \mathcal{D}J^* \exp \left[ -\frac{1}{\sigma^2} \sum_{ijkl} \left( J_{ijkl} J_{ijkl}^* + \frac{1}{2} J_{ijkl} \int d\tau \ c_{im}^\dagger c_{jm}^\dagger c_{km} c_{lm} + \left( \frac{1}{2} J_{ijkl} \int d\tau \ c_{im}^\dagger c_{jm}^\dagger c_{km} c_{lm} \right)^\dagger \right) \right] \nonumber \\
=&& \int \mathcal{D}J \mathcal{D}J^* \exp \left[ -\frac{1}{\sigma^2} \sum_{ijkl} \left| J_{ijkl} + \frac{\sigma^2}{2} J_{ijkl} \int d\tau \ c_{im}^\dagger c_{jm}^\dagger c_{km} c_{lm} \right|^2 \right] \exp \left[ \frac{\sigma^2}{4} \sum_{ijkl} \left| \int d\tau \ c_{im}^\dagger c_{jm}^\dagger c_{km} c_{lm} \right|^2 \right] \nonumber \\
=&& \exp \left[ \frac{\sigma^2}{4} \sum_{ijkl} \int d\tau d\tau^\prime \ c_{im}^\dagger (\tau) c_{jm}^\dagger (\tau) c_{km} (\tau) c_{lm} (\tau) c_{ln}^\dagger (\tau^\prime) c_{kn}^\dagger (\tau^\prime) c_{jn} (\tau^\prime) c_{in} (\tau^\prime) \right],
\eea
where \( \mathcal{D}J \mathcal{D}J^* \) is shorthand for \( \prod_{ijkl} dJ_{ijkl} dJ_{ijkl}^* \), and we use the condition \( J_{ijkl} = J_{klij}^* \).
By employing the anticommutation relations, we can rearrange the fermion operators in the Gaussian integral (\ref{gaussian integral}). This manipulation leads to
\bea
&& \int \mathcal{D}J \mathcal{D}J^* \exp \left[ -\sum_{ijkl} \frac{|J_{ijkl}|^2}{\sigma^2} \right] \exp \left[ -\sum_{ijkl} J_{ijkl} \int d\tau \ c_{im}^\dagger c_{jm}^\dagger c_{km} c_{lm} \right] \nonumber \\
&=& \exp \left[ \frac{\sigma^2}{4} \int d\tau d\tau^\prime \left( \sum_i c_{im}^\dagger (\tau) c_{in} (\tau^\prime) \right)^2 \left( \sum_i c_{im} (\tau) c_{in}^\dagger (\tau^\prime) \right)^2 \right].
\eea
Upon integrating out the Gaussian random variables, the resulting expression for the averaged partition function is
\bea
\langle Z^M \rangle_J &=& \int \prod_{i,m} \mathcal{D}c_{im} \mathcal{D}c_{im}^\dagger \exp \left[ -\sum_{m=1}^M \int d\tau \sum_i c_{im}^\dagger (\partial_\tau - \mu) c_{im} \right. \nonumber \\
&& \left. + \sum_{m,n=1}^M \frac{\sigma^2}{4} \int d\tau d\tau^\prime \left( \sum_i c_{im}^\dagger (\tau) c_{in} (\tau^\prime) \right)^2 \left( \sum_i c_{im} (\tau) c_{in}^\dagger (\tau^\prime) \right)^2 \right].
\eea
Here, the integration over the Gaussian random variables has simplified the partition function significantly, making it more amenable to further analysis and manipulation. This approach reduces the complexity inherent in the original formulation by leveraging the anticommutation properties of the fermion operators.

Moving forward, we can introduce the following identity
\bea\label{Multiplier}
1 &=& \int \mathcal{D}G~\prod_{m,n} \delta\left(G_{mn}(\tau,\tau^\prime) - \frac{1}{N} \sum_i c_{im}^\dagger(\tau) c_{in}(\tau^\prime) \right) \nonumber\\
&=& \int \mathcal{D}G \mathcal{D}\Sigma \exp \left[ N \sum_{m,n} \int d\tau d\tau^\prime \Sigma_{mn}(\tau,\tau^\prime) \left(G_{mn}(\tau,\tau^\prime) - \frac{1}{N} \sum_i c_{im}^\dagger(\tau) c_{in}(\tau^\prime) \right) \right],
\eea
where the second equality in (\ref{Multiplier}) is derived using the exponential representation of the delta function. Here, the Lagrange multiplier \( \Sigma_{mn} \) acts as the fermionic self-energy of the system. The delta function satisfies the relation
\bea
f\left(\frac{1}{N} \sum_i c_{im}^\dagger(\tau) c_{in}(\tau^\prime)\right) = \int \mathcal{D}G \ f\left(G_{mn}(\tau,\tau^\prime)\right) \delta\left(G_{mn}(\tau,\tau^\prime) - \frac{1}{N} \sum_i c_{im}^\dagger(\tau) c_{in}(\tau^\prime) \right).
\eea
By substituting the identity (\ref{Multiplier}) into the replicated partition function, we obtain
\bea
\langle Z^M \rangle_J &=& \int \prod_{i} \mathcal{D}c_i^\dagger \mathcal{D}c_i \mathcal{D}G \mathcal{D}\Sigma \ \exp \left\{
- \sum_{i,m} \int d\tau \ c_{im}(\tau)(\partial_{\tau} - \mu) c_{im}(\tau) \right. \nonumber\\
&-& N \sum_{m,n} \int d\tau d\tau^\prime \ c_{nm}^\dagger(\tau) \Sigma_{nm}(\tau,\tau^\prime) c_{nm}(\tau^\prime) \nonumber\\
&+& N \sum_{m,n} \int d\tau d\tau^\prime \ \Sigma_{nm}(\tau,\tau^\prime) G_{nm}(\tau,\tau^\prime) \nonumber\\
&+& \left. \frac{1}{4} \int d\tau d\tau^\prime \ G_{nm}^2(\tau,\tau^\prime) G_{nm}^2(\tau^\prime,\tau) \right\}.
\eea
This expression integrates out the fermionic degrees of freedom, leaving a functional integral over \( G \) and \( \Sigma \). To streamline our analysis, we can assume the Green's functions \( G_{nm} \) and the self-energy \( \Sigma_{nm} \) are diagonal. This means that all off-diagonal terms where \( m \neq n \) are negligible, resulting in \( \Sigma_{mn}(\tau,\tau^\prime) = \delta_{mn} \Sigma(\tau,\tau^\prime) \) and \( G_{mn}(\tau,\tau^\prime) = \delta_{mn} G(\tau,\tau^\prime) \).

After averaging over the disorder and retaining only the replica-diagonal components, we integrate out the fermion fields. This process gives us the effective action \( S_{\rm eff} = -\ln \langle Z \rangle_J \) in the form
\bea
-\frac{S_{\rm eff}}{N} &=& \ln \det \left[ (\partial_{\tau} - \mu) - \Sigma \right] + \int d\tau d\tau' \left[ \Sigma(\tau, \tau') G(\tau', \tau) + \frac{J^2}{4} G^2(\tau, \tau') G^2(\tau', \tau) \right].
\eea
The auxiliary fields \( G(\tau, \tau^\prime) \) and \( \Sigma(\tau, \tau^\prime) \) are instrumental in applying the saddle point method. For the effective action's saddle point, the conditions \( \frac{\delta S_{\rm eff}}{\delta G} = 0 \) and \( \frac{\delta S_{\rm eff}}{\delta \Sigma} = 0 \) must be met, yielding the Schwinger-Dyson equations
\bea \label{SD eq}
G(i\omega_n) &=& \frac{1}{-i\omega_n - \mu - \Sigma}, \nonumber\\
\Sigma(\tau) &=& -J^2 G^2(\tau) G(-\tau),
\eea
where \( \omega_n = (2n + 1) \frac{\pi}{\beta} \) represent Matsubara frequencies. These Schwinger-Dyson equations are crucial for comprehending the model's thermodynamic properties. By solving them numerically, we can explore the model's behavior beyond the low-energy regime.

Physically, the Schwinger-Dyson equations encapsulate the interplay between the Green's function and the self-energy in a strongly correlated system. The first equation describes the Green's function in the presence of self-energy corrections, which account for the interactions between fermions mediated by the random couplings \( J_{ijkl} \). The second equation reflects the feedback of the Green's function on the self-energy, highlighting the non-trivial dynamics introduced by the interaction term. The temperature dependence embedded in the Matsubara frequencies \( \omega_n \) allows us to probe different regimes of the model. At high temperatures, thermal fluctuations dominate, leading to a different behavior compared to the low-temperature, quantum-dominated regime. 
The Schwinger-Dyson equations serve as a powerful framework for understanding the complex interactions in the SYK model, offering a window into the rich physics of disordered, strongly interacting systems.
\subsection{Particle-hole symmetry breaking and phase transition}
In the case of the cSYK model, when the chemical potential is zero, the system exhibits particle-hole symmetry \cite{1910.14099,Fu_2016,Ferrari}. This particle-hole symmetry leads to a key feature that the imaginary-time Green's function is symmetric about \(\beta/2\), i.e.,

\bea
G(\tau + \beta/2) = G(-\tau + \beta/2)\quad {\rm or~equivalently}\quad G(\tau ) = -G(-\tau),
\eea
which is identical to the behavior observed in the real SYK model constructed from Majorana fermions. In this neutral case, the complex SYK model and the real SYK model share several key characteristics, including the same form of the SD equations. Additionally, both models do not exhibit any thermodynamic phase transitions.

However, when the chemical potential is introduced, the corresponding term reads
\bea
H_{\mu} = -\mu \sum_i c_i^\dagger c_i
\eea
explicitly breaks the particle-hole symmetry. Under the particle-hole symmetry transformation (\(\psi \to \psi^\dagger\)), this term transforms as
\bea
H_{\mu} \to -\mu N-H_{\mu}.\nonumber
\eea
This breaking of the symmetry has a direct impact on the Green's function. In the presence of a nonzero chemical potential, the imaginary-time Green's function no longer maintains the symmetry about \(\beta/2\). This symmetry breaking is reflected in the Matsubara frequency space, where the Green's function \( G(i\omega_n) \) is no longer anti-symmetric. Specifically, the real part of \( G(i\omega_n) \) becomes a symmetric function, while the imaginary part remains anti-symmetric. Thus, the Green's function takes on a more general form, with both symmetric and anti-symmetric components, breaking the particle-hole symmetry in such behavior.

This explicit breaking of symmetry, caused by the chemical potential, directly leads a redistribution of the energy spectrum density. In section \ref{ED}, we study in detail how the introduction of the chemical potential term affects the distribution of the energy spectrum by using the exact diagonalization numerical method in the finite $N$ case. It also causes a shift in the spectral properties of the Green's function (see section \ref{realtime} for details), reflecting a fundamental difference between the neutral cSYK model and the one with a nonzero chemical potential.

The presence of a chemical potential modifies the distribution of fermionic occupancy and hole states in the system. The explicit breaking of symmetry has a direct impact on the solution space of the system, which can be observed in the SD equations. In particular, the SD equations no longer have a single solution; instead, for certain temperatures and chemical potentials, the system is allowed to converge to multiple distinct solutions, where the free energy of the system typically exhibits multiple stable solutions. The presence of these multiple solutions is a signature of a first-order phase transition. To understand this, consider the stability of the solutions: below a certain critical temperature $T_c$, multiple solutions become stable, and the system can select between different physical states. As the temperature varies, the system may transit from one state to another, leading to a first-order phase transition.

The transition from non-physical to physical solutions in the SD equations is not clearly reflected in the $q=4$ case, where only numerical solutions are available. To illustrate this, we turn to the $q=2$ case, where the SD equations can be solved analytically. The detailed discussion can be found in Appendix \ref{appendix1}. The Green's function for $q=2$ cSYK model in Matsubara frequency reads,
\bea
G(i\omega_n)=\frac{-i\omega_n-\mu+\sqrt{(i\omega_n+\mu)^2-4J^2}}{2J^2}.
\eea
The conclusion is that, with the introduction of the chemical potential, we indeed observe that the Green's function, which originally diverged at infinity in Matsubara frequency space, immediately becomes convergent. Moreover, the numerical solution can also converge to this branch. As previously mentioned, the Green's function no longer maintains the original odd symmetry, indicating that the chemical potential term, which breaks the symmetry, enables the SD equations to possess more acceptable solutions. The Green's function, once constrained by particle-hole symmetry, is no longer bound and becomes both convergent and physically available since the properties of particle-hole symmetry no longer constrain the behavior of the Green's function.

In this context, the appearance of multiple solutions in the SD equations, corresponding to different states of the system, reflects the breaking of the particle-hole symmetry and indicates a transition from one phase to another as the temperature decreases. This transition is a hallmark of a first-order phase transition, characterized by a discontinuous change in free energy and a clear switching between stable states.

The explicit breaking of particle-hole symmetry due to the chemical potential plays a crucial role in facilitating this transition, causing the system to exhibit multiple stable states that are energetically distinct. This phenomenon underscores the close relationship between symmetry breaking and phase transitions in many-body systems.

\subsection{Numerical solution of Schwinger-Dyson equations }
Once we have derived the Schwinger-Dyson equations, the primary challenge lies in solving them efficiently. In the infrared limit, if we neglect the term \(-i\omega_n\) in the saddle point equation, we can analytically derive the conformal symmetrical solution for the Green's function. However, our primary interest is in the thermodynamic properties and critical phenomena of the model, which necessitates exploring beyond the low-energy limit. Therefore, numerical calculations become essential for our analysis.

The effective action is given by
\bea
-\frac{S_{\rm eff}}{N} = \ln {\rm det} \bigg[(\partial_{\tau} - \mu) - \Sigma\bigg] + \int d\tau d\tau' \bigg[\Sigma (\tau, \tau') G (\tau', \tau) + \frac{J^2}{4} G^2 (\tau, \tau') G^2 (\tau', \tau) \bigg].
\eea
Owing to the time translation invariance of Euclidean correlators, \( G(\tau, \tau^\prime) = G(\tau - \tau^\prime) \), we can reformulate the action as
\bea
-\frac{S_{\rm eff}}{N} = \ln {\rm det} \bigg[\partial_{\tau} - \mu - \Sigma\bigg] + \beta \int d\tau \bigg[\Sigma (\tau) G (-\tau) + \frac{J^2}{4} G^2 (\tau) G^2 (-\tau) \bigg].
\eea
Utilizing the Fourier transformation,
\bea
f(\tau) = \frac{1}{\beta} \sum_{\omega_n} e^{-i \omega_n \tau} f(i\omega_n), \quad f(i\omega_n) = \int_{0}^{\beta} d\tau e^{i \omega_n \tau} f(\tau),
\eea
we obtain the effective action in terms of frequency components
\bea\label{Seff}
-\frac{S_{\rm eff}}{N} = \ln 2 + \sum_{\omega_n} \frac{1}{2} \ln \left( \frac{ -i \omega_n - \mu - \Sigma(i \omega_n)}{-i \omega_n} \right)^2 + \frac{3}{4} \sum_{\omega_n} \Sigma(i \omega_n) G(i \omega_n).
\eea
To solve the Schwinger-Dyson equations derived previously, we need to find a fixed-point equation of the form \( G(i \omega_n) = f \left(G(i \omega_n)\right) \). We use the algorithm introduced in \cite{stanford} to perform numerical calculations for this purpose. Starting with an initial guess for the correlator, the algorithm iteratively updates the solution
\bea
G^{n+1}(i \omega_n) = (1 - x) G^{n}(i \omega_n) + x \frac{1}{-i \omega_n - \mu - \Sigma(i \omega_n)},
\eea
where we set the update parameter \( x = 0.05 \). The superscript on the Green's functions denotes the iteration step. Initially, we assume the noninteracting solution \( G(\tau) = \frac{1}{2} {\rm sgn}(\tau) \) for the first iteration, with the coupling strength \( J \) set to one. For numerical implementation \cite{1912-12302}, we discretize Euclidean time and Matsubara frequencies as
\bea
&&\tau_m = \frac{\beta m}{2 \Lambda}, \quad (m = 0, 1, \dots, 2 \Lambda - 1), \nonumber\\
&&\omega_n = \frac{2 \pi}{\beta} \left( n + \frac{1}{2} \right), \quad (n = -\Lambda, -\Lambda + 1, \dots, \Lambda - 1),
\eea
introducing a UV cutoff. Typically, we choose \(\Lambda = 2^{16}\) for accuracy. After the final iteration, we evaluate the convergence by computing the difference
\bea
\Delta G = \frac{1}{\Lambda} \sum_{\omega_n} \left| G^{n+1}(i \omega_n) - G^{n}(i \omega_n) \right|.
\eea
When the difference is below an acceptable limit $\Delta G<\epsilon$, in which we set $\epsilon=10^{-14}$ as our threshold, the algorithm terminates. This indicates that the iterative solution is sufficiently close to the true solution. We thus can argue the numerical solution for Green function at this moment is converged. It is noteworthy that if $G^{n}(i\omega_n)$ is the real solution that satisfy the Schwinger-Dyson equations (\ref{SD eq}), then the next iteration will yield the same solution $G^{n+1}(i\omega_n)=G^{n}(i\omega_n)$, resulting in $\Delta G=0$.
\begin{figure}[t]
	\center{
		\includegraphics[height=6cm, width=8.8cm]{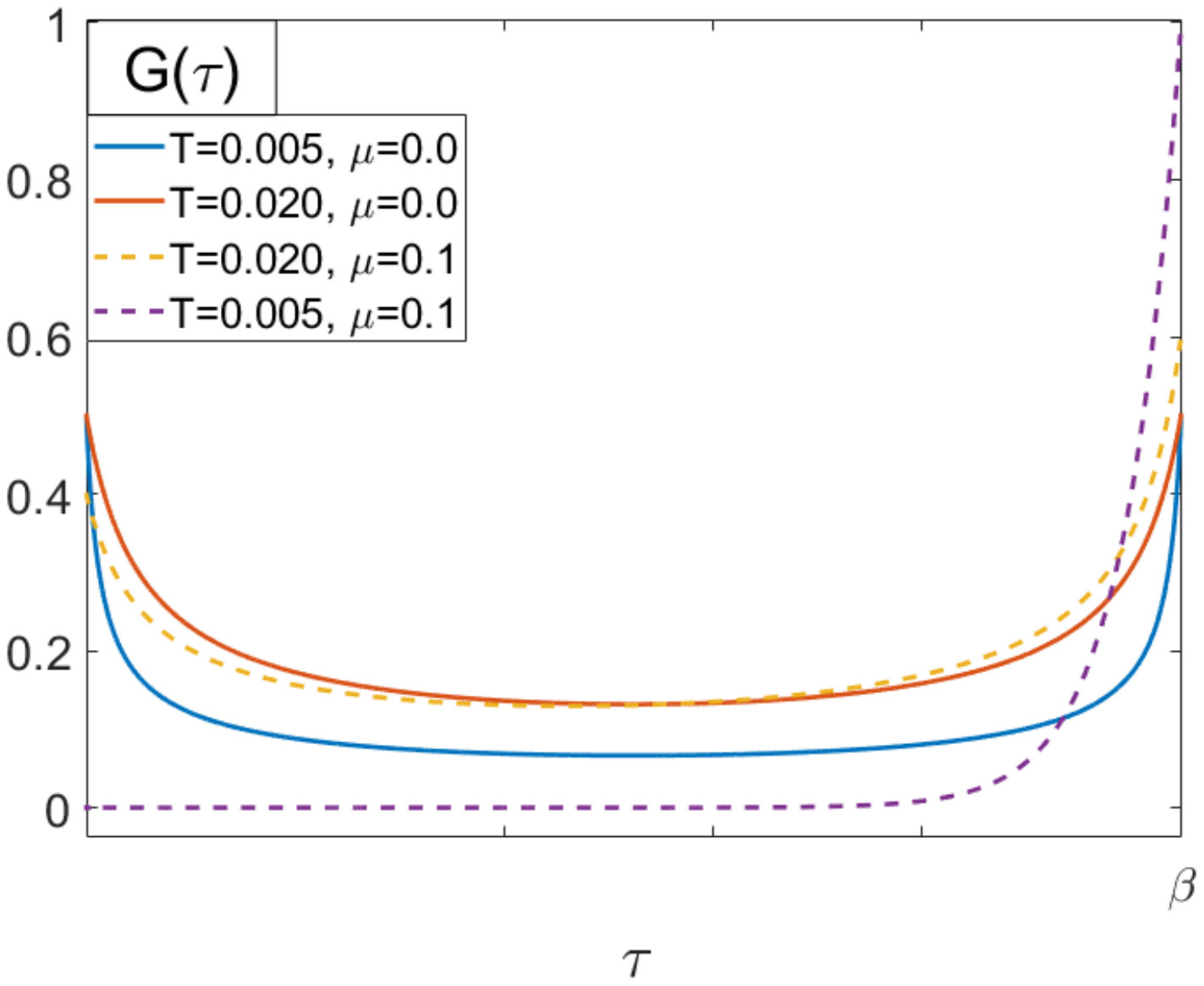}
		\caption{Numerical solution of $G(\tau)$ for different temperatures and chemical potentials. The calculations are done for $J=1$ and $q=4$ and $\Lambda=2\times 10^{17}$.}\label{Gtau0}}
\end{figure}

Fig. (\ref{Gtau0}) shows the numerical solution of the Green function $G(\tau)$ at different temperatures and chemical potentials. One can observe that the numerical solution of $G(0^{+})+G(\beta^{-})=1$ holds true, and a gap appears at low temperatures in the presence of a chemical potential $\mu$.

\subsection{Thermodynamic properties of the cSYK}
After obtaining the numerical solution for the Euclidean Green's function, we can express the system's energy as a functional of this Green's function. By following the approach outlined in \cite{MQ18}, we derive the energy expression. To begin, we consider the following relationship
\bea\label{commutator}
N\partial_{\tau_1} G(\tau_1, \tau_2)\big{|}_{\tau_1 - \tau_2 = 0^+} &=& N\partial_{\tau_1} G(\tau_1 - \tau_2)\big{|}_{\tau_1 - \tau_2 = 0^+} \nonumber\\
&=& \sum_{i} \partial_{\tau} \langle c_{i}(\tau)c_{i}^\dagger \rangle = \sum_{m} \langle [H, c_{m}]c_{m}^\dagger \rangle,
\eea
where the Hamiltonian \(H\) is divided into two components
\bea
&&H = H_{\rm SYK} + H_{0}, \\
&&H_{\rm SYK} = \sum J_{ijkl} c_{i}^\dagger c_{j}^\dagger c_{k} c_{l}, \\
\label{H0}
&&H_0 = -\mu \sum_{i} c_{i}^\dagger c_{i}.
\eea
Substituting the commutator \([H, c_{m}]\) back into the earlier equation (\ref{commutator}), we obtain
\bea
N\partial_{\tau} G(\tau)\big{|}_{\tau = 0^+} = \langle qH_{\rm SYK} + H_{0} \rangle.
\eea
Incorporating additional terms, we derive the relationship between the Green's functions and the energy
\bea
N\partial_{\tau} G(\tau)\big{|}_{\tau = 0^+} - \mu(q - 1)\langle Q + \frac{N}{2} \rangle = \langle qH_{\rm SYK} + H_{0} + (q - 1)H_{0} \rangle = q\langle H \rangle,
\eea
which implies that the average energy per particle is given by
\bea\label{E/N}
\frac{E}{N} = \frac{1}{q} \left[ \partial_{\tau} G(\tau)\big{|}_{\tau = 0^+} - \mu(q - 1)\left(\mathcal{Q} + \frac{1}{2}\right) \right],
\eea
where \(\mathcal{Q}\) is the \(U(1)\) conserved charge density, defined as
\bea
\mathcal{Q} \equiv \frac{Q}{N} = -\frac{1}{2} \left[G(0^+) + G(0^-)\right].
\eea

To numerically evaluate the energy, it is crucial to compute the first term in the energy expression (\ref{E/N}). The saddle-point equation in our model is
\bea
(\partial_\tau - \mu) G(\tau) - \int_{0}^{\beta} \Sigma(\tau - \tilde{\tau}) G(\tilde{\tau}) d\tilde{\tau} = -\delta_\beta(\tau),
\eea
which can be rewritten as
\bea
\partial_\tau G(\tau) = \mu G(\tau) + \int_{0}^{\beta} \Sigma(\tau - \tilde{\tau}) G(\tilde{\tau}) d\tilde{\tau} + \delta_\beta(\tau).
\eea
Taking the limit \(\tau \rightarrow 0^+\) and using the expression for \(\mathcal{Q}\), we obtain
\bea
\partial_{\tau} G(\tau)\big{|}_{\tau = 0^+} = \mu \left(\frac{1}{2} - \mathcal{Q}\right) + \frac{1}{\beta} \sum_{\omega_n} \Sigma(i\omega_n) G(i\omega_n).
\eea

This formulation allows us to numerically compute the energy by evaluating the Green's function and self-energy at specific points. Accurate numerical treatment ensures precise results, particularly near critical points where the system's behavior is highly sensitive to parameter changes.

\subsection{Deriving the energy functional}
For the numerical evaluation of the energy, the first term in equation (\ref{E/N}) must be computed. The saddle-point equation for our model is
\bea
(\partial_\tau - \mu) G(\tau) - \int_{0}^{\beta} \Sigma(\tau - \tilde{\tau}) G(\tilde{\tau}) d\tilde{\tau} = -\delta_\beta(\tau),
\eea
which implies
\bea
\partial_\tau G(\tau) = \mu G(\tau) + \int_{0}^{\beta} \Sigma(\tau - \tilde{\tau}) G(\tilde{\tau}) d\tilde{\tau} + \delta_\beta(\tau).
\eea
Taking the limit \(\tau \rightarrow 0^+\) and using the expression for \(\mathcal{Q}\), we derive
\bea
\partial_{\tau} G(\tau)\big{|}_{\tau = 0^+} = \mu \left(\frac{1}{2} - \mathcal{Q}\right) + \frac{1}{\beta} \sum_{\omega_n} \Sigma(i\omega_n) G(i\omega_n).
\eea
Thus, the computable formula for the energy becomes
\bea
\frac{E}{N}=\frac{1}{q} \big[4\mu T\sum_{\omega_n}{\rm Re}[G(i\omega_n)]-\mu+T\sum_{\omega_n}\Sigma(i\omega_n)G(i\omega_n)\big],\nonumber\\
\eea
where we used the Dirichlet trick to ensure that the series converges
\bea
\mathcal{Q}=-\frac{1}{2}[G(0^+)+G(0^-)]=-T\sum_{\omega_n}Re[G(i\omega_n)].
\eea
This framework enables the numerical computation of the energy by evaluating the Green's function and self-energy at specific points. Precise numerical treatment is crucial to ensure accurate results, especially near critical points where the system's behavior is highly sensitive to changes in parameters. Substituting the saddle point solution in (\ref{Seff}), we obtain the free energy
\bea
\frac{F}{N}=-T\bigg[\ln2&+&\sum_{\omega_n} \frac{1}{2}\ln \left( \frac{ -i\omega_n-\mu-\Sigma(i\omega_n)}{-i\omega_n} \right)^2\nonumber\\&+&\frac{3}{4}\sum_{\omega_n}\Sigma(i\omega_n)G(i\omega_n)\bigg].
\eea
\begin{figure}[t]
	\subfigure[]{\label{be1}
		\includegraphics[height=4.0cm, width=4.3cm]{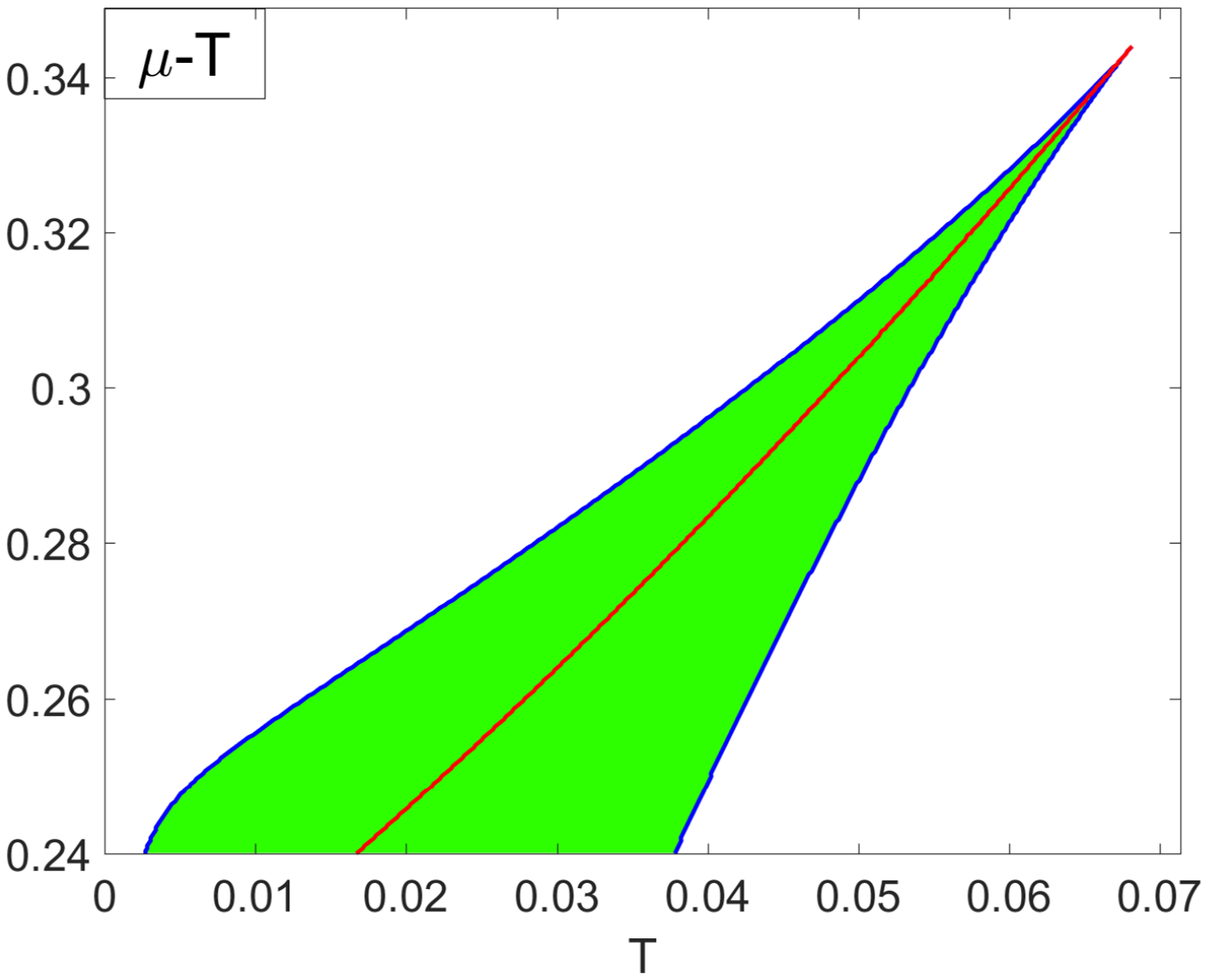}}
	\subfigure[]{\label{FT03}
		\includegraphics[height=4.0cm, width=4.3cm]{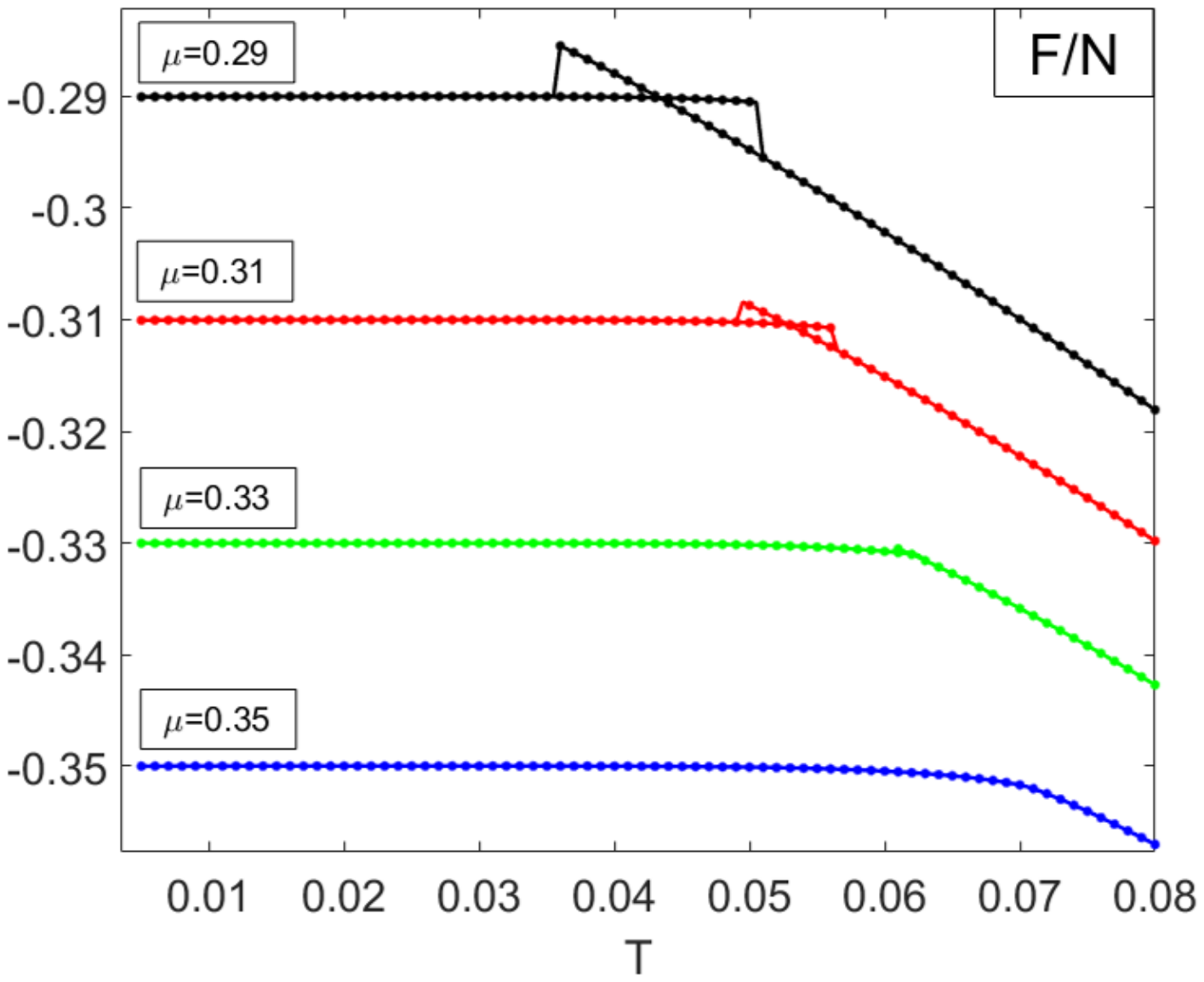}}
	\subfigure[]{\label{Fmu}
		\includegraphics[height=4.0cm, width=4.3cm]{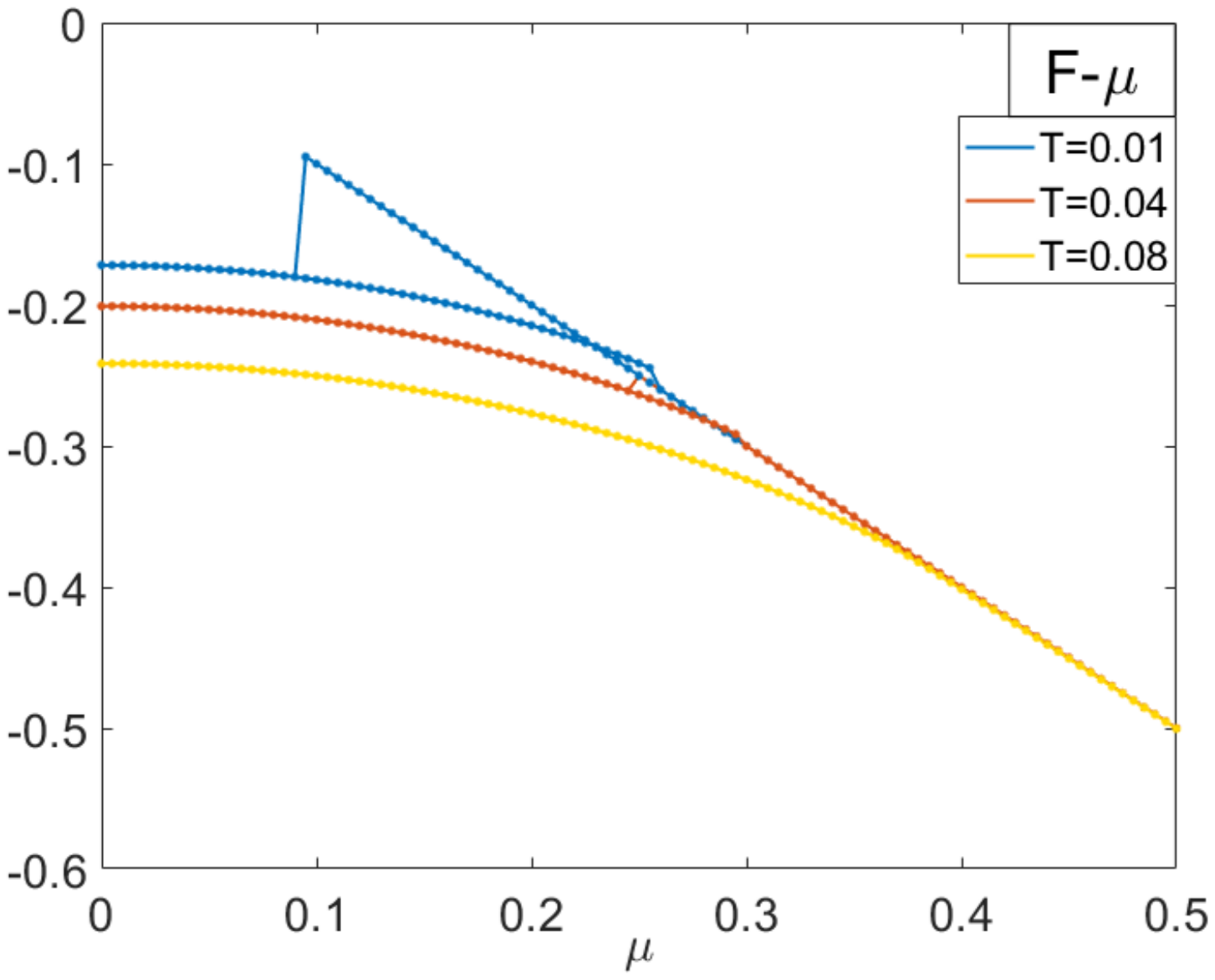}}
	\subfigure[]{\label{QT}
		\includegraphics[height=4.3cm, width=4.2cm]{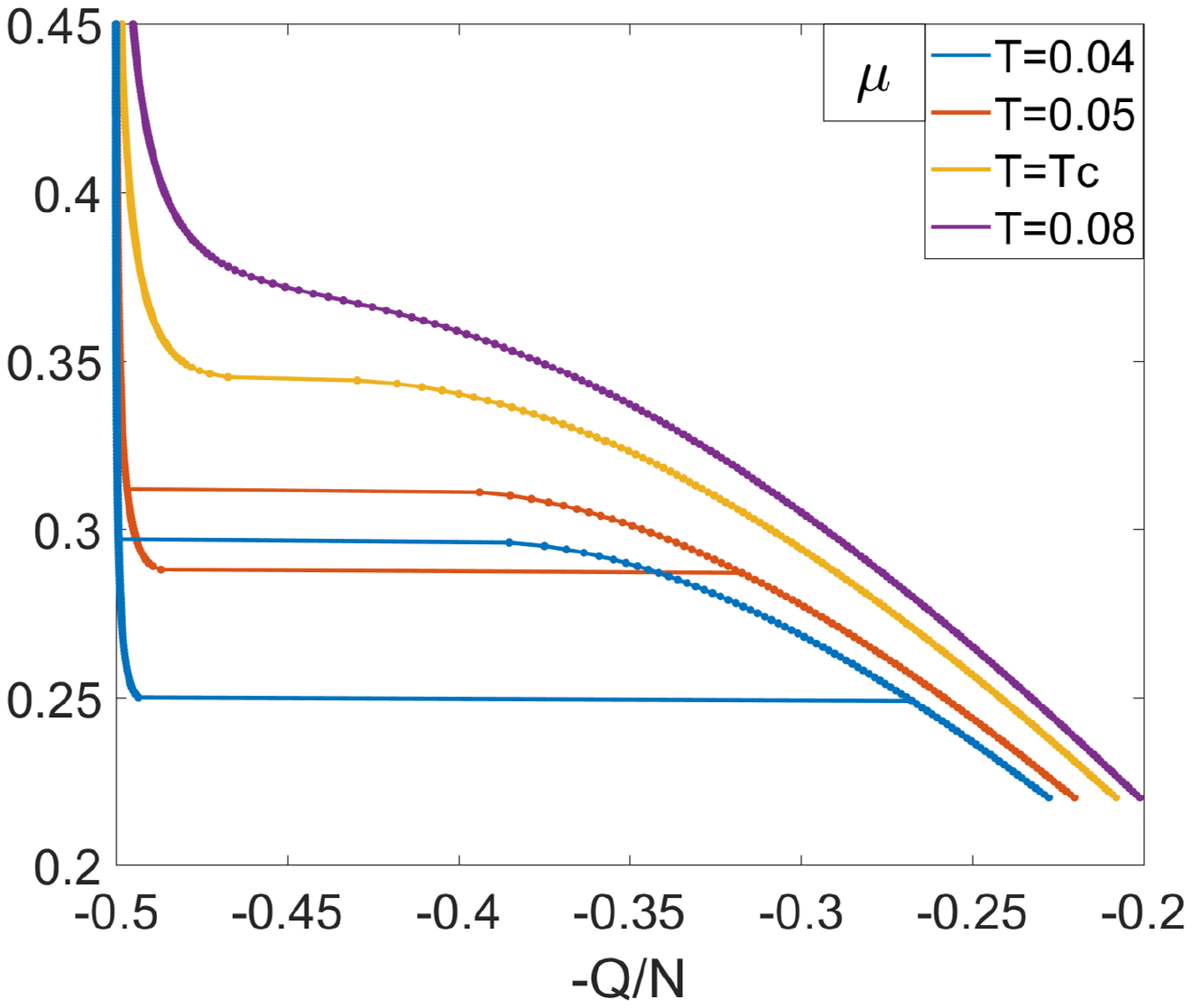}}

	\caption{ (a) The chemical potential as a function of $T$. The shaded region indicates the parameter range where the phases coexist.
		(b) The free energy as a function of temperature for different  $\mu$. 
		(c) The free energy as a function of chemical potential for different temperatures. 
		(d) The chemical potential as a function of the charge density for different $T$.
	}\label{fig2}
\end{figure}
The key numerical results are presented in Fig. (\ref{fig2}a-d). The phase diagram of the single-sided cSYK model features a critical point at \((T_c, \mu_c) = (0.06828, 0.3443)\), as shown in Fig. (\ref{be1}). High temperatures with fixed \(\mu\) and high \(\mu\) with fixed \(T\) in the \((\mu, T)\)-plane correspond to gapless, strongly correlated SYK phases and gapped, weakly coupled fermion phases, respectively \cite{aze17, Ferrari}.

The system behaves like an insulator with weakly interacting fermions described by perturbative field theory in the large \(\mu\) regime. In the small \(\mu\) regime, it represents a strongly correlated metallic phase, akin to a black hole. Adjusting the chemical potential allows us to explore the transition between weakly and strongly coupled regimes.

The transition temperature \(T_c\) increases with \(\mu\) and culminates at the critical point \((T_c, \mu_c)\). At higher temperatures, the black hole-like metallic phase dominates, while at lower temperatures, the weakly coupled fermionic phase is more stable. Hysteresis between these phases, shown in Fig (\ref{FT03}), indicates a first-order phase transition ending at the critical point. Beyond a critical \(\mu\), hysteresis disappears, and the free energy becomes a smooth function of temperature.

Fig (\ref{Fmu}) illustrates the system's free energy as a function of \(\mu\). As \(\mu\) nears the critical value \(\mu_c\), the first-order phase transition occurs at higher temperatures, eventually smoothing out the kink in the free energy. At high \(\mu\), a single saddle point solution exists, indicating the weakly coupled fermion phase's dominance.

These findings align with the \((\mu, -\mathcal{Q})\) diagram in Fig. (\ref{QT}), where the charge density difference acts as an order parameter. Below \(T_c\), certain \(\mathcal{Q}\) ranges (non-dotted parts of the line) lack saddle-point solutions, even in the large \(q\) limit. These missing segments indicate thermodynamically unstable branches with negative specific heat. Similar behavior of charge density as a function of chemical potential is also observed in Fig. (11a) of \cite{Tikhanovskaya_2021}, where they employ a different numerical method called conformal perturbation theory to study low-energy behaviour.

\begin{figure}[t]
    \centering
    \subfigure[]{\label{G-P_vdWs}
        \includegraphics[height=5cm, width=7.0cm]{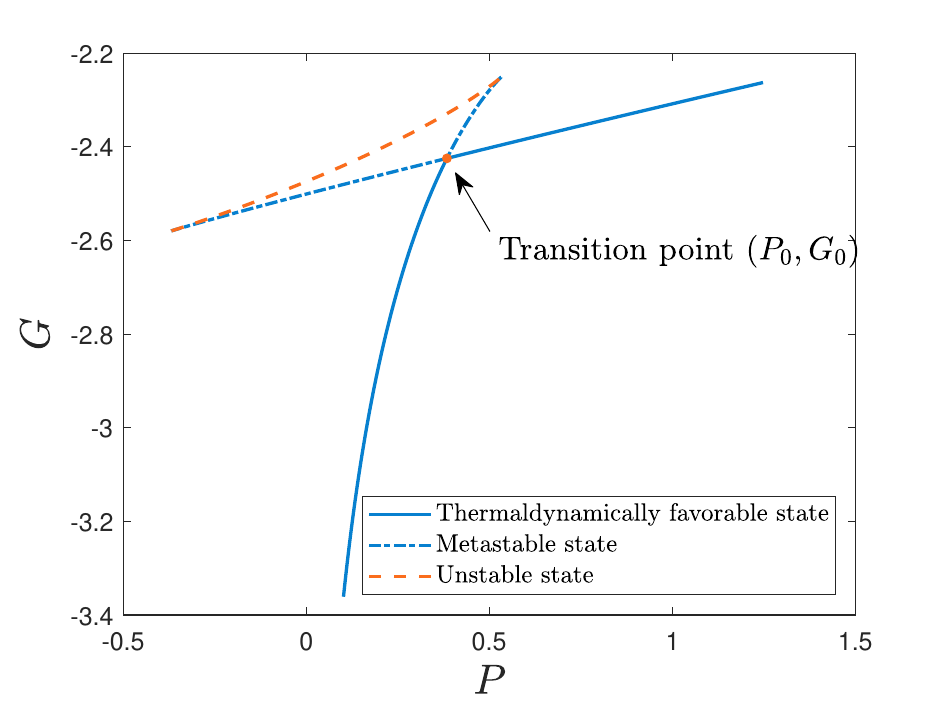}}
    \subfigure[]{\label{P-V_vdWs}
        \includegraphics[height=5cm, width=7.0cm]{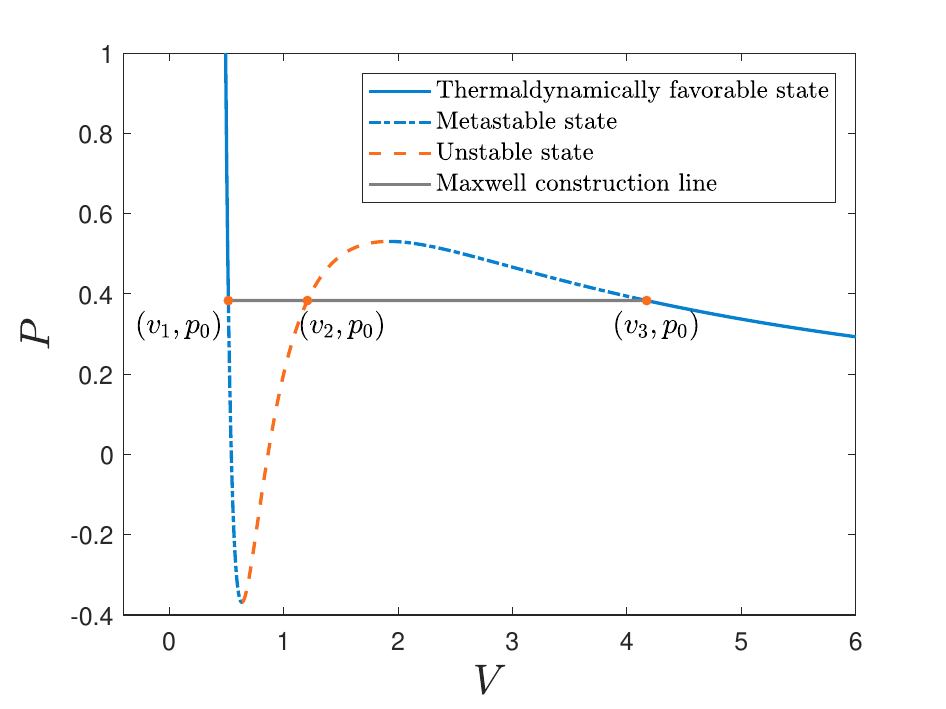}}
    \subfigure[]{\label{F-mu_SYK}
        \includegraphics[height=5cm, width=7.0cm]{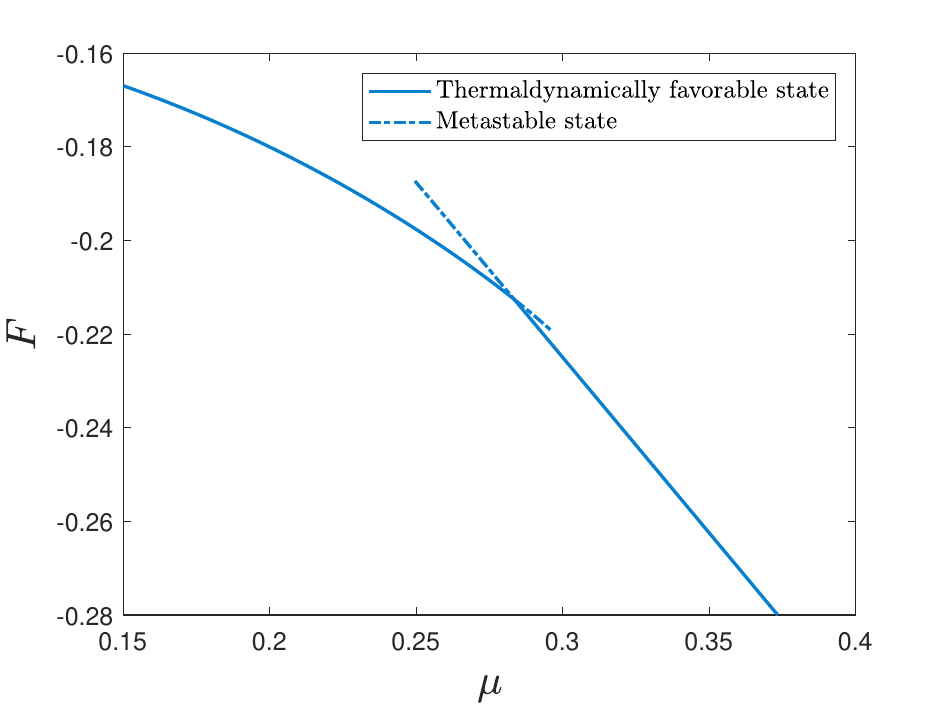}}
    \subfigure[]{\label{mu-Q_SYK}
        \includegraphics[height=5cm, width=7.0cm]{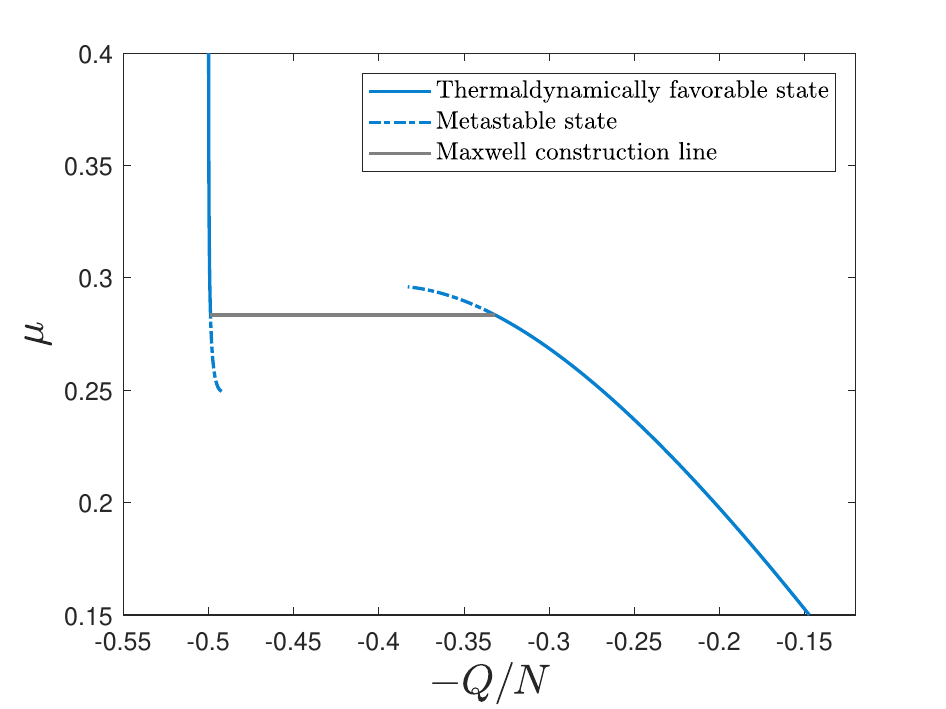}}
    \caption{(a) The Gibbs free energy $G$ as a function of pressure $P$ for a van der Waals gas at $T=0.8T_c$. (b) $P-V$ diagram of isotherms at the same temperature. (c) The free energy $F$ as a function of chemical potential $\mu$ for cSYK model at $T=0.04$. (d) The corresponding Maxwell construction line in the same isotherm for cSYK model.}\label{cSYK-vdWs}
\end{figure}

We compare the phase transition of cSYK model with the van der Waals phase transition, and find that the chemical potential $\mu$ in the cSYK model behaves like the pressure $P$ in the van der Waals gas, while the charge density $Q/N$ behaves the same as the volume behavior of the van der Waals gas. The van der Waals equation with critical temperature $T_c=1$ reads
\bea
P=\frac{8T}{3V-1}-\frac{3}{V^2}.
\eea
Fig. (\ref{cSYK-vdWs}a-d) illustrates how the SYK model is similar to the van der Waals gas model in several respects. Both the van der Waals equation and the solutions to the SD equations exhibit multivalued phenomena when the temperature falls below the critical temperature, where a single volume corresponds to multiple pressures or a single chemical potential corresponds to multiple charge densities. As shown in Fig. (\ref{G-P_vdWs}), within the phase transition region, the solution with the lowest free energy is thermodynamically favorable, the second lowest corresponds to a metastable solution, and the highest free energy solution is unstable, corresponding to a negative heat capacity at that branch. During the gas-liquid phase transition process, gas and liquid are in equilibrium at a specific pressure, implying that their volume and pressure should be monotonic at that temperature. 

Maxwell's equal area law corrects this unreasonable phenomenon in the phase transition process by balancing conditions, aligning the mathematical model of the liquid-gas phase transition with actual physical phenomena. More specifically, in the van der Waals $P$-$V$ relation Fig. ($\ref{P-V_vdWs})$, the unstable region determined by Maxwell's equal area law corresponds to a triangular area on the corresponding $G$-$P$ graph, where the thermodynamically favorable solution always corresponds to the solution that minimizes the free energy. Similarly, such behaviors are observed in the solutions to the SD equations of SYK, where two types of solutions can appear at the same chemical potential or temperature within the phase transition region, corresponding to different free energies, energies, charge densities, and entropies. As generally described by the van der Waals gas, the thermodynamically favorable solution should be those that minimize the free energy.

Non-physical cases are indicated with dashed lines, and it can be observed that the behavior of the van der Waals gas's $G$-$P$ graph is similar to the SYK's $F$-$\mu$ graph as shown in Fig. (\ref{F-mu_SYK}): both exhibit a non-analytic point with a discontinuous first derivative linking the transition between two phases, representing a first-order phase transition. Before and after the phase transition, the free energy transitions from nonlinear to linear behavior, with the van der Waals gas transitioning from gas to liquid phase, and the SYK model transitioning from a gapped to a gapless phase. Due to numerical methods, for a given chemical potential, the SD equation converges to at most two solutions, stable and metastable, while unstable state solutions do not converge stably due to algorithm characteristics and will eventually converge to either a stable or a metastable solution, depending on which is closer. This can also be verified using the van der Waals equation. First rewrite the van der Waals equation as a fixed point equation, i.e. \bea\label{vdW-fix-point}
V=g(V)=(8TV^2-(3V-1)V^2P+3)/9.
\eea
Given $P=p_0$, using the fixed-point iteration method employed for solving the SD equations similarly fails to find the volumes corresponding to unstable states $(v_2, p_0)$. The criterion for whether the fixed-point iteration converges to a solution is whether the absolute value of the derivative of $g(V)$ at a solution, denoted as $|g'(V)|$, is less than 1, which is consistent with the fact that the heat capacity of the unstable solution is negative. However, inspired by Maxwell's construction law, we can still determine the chemical potential plateau that satisfies the area law at the intersection point of the two branches in the $F$-$\mu$ graph, as shown by the grey line in Fig. (\ref{mu-Q_SYK}).

In principle, for a cubic equation like \eqref{vdW-fix-point}, three complex roots can always exist. The number of physically acceptable real roots is then determined by the discriminant, which depends on the coefficients (and in this case, on temperature and pressure). From this perspective, the role of the chemical potential is analogous to that of pressure. By adjusting the chemical potential and temperature of the cSYK model, multiple physical solutions to the SD equations can be obtained. That is, the values of chemical potential and temperature determine which of the theoretically possible multi-branch solutions are physically acceptable.

\begin{figure}[t]
	\subfigure[]{\label{E-T}
		\includegraphics[height=5cm, width=7.2cm]{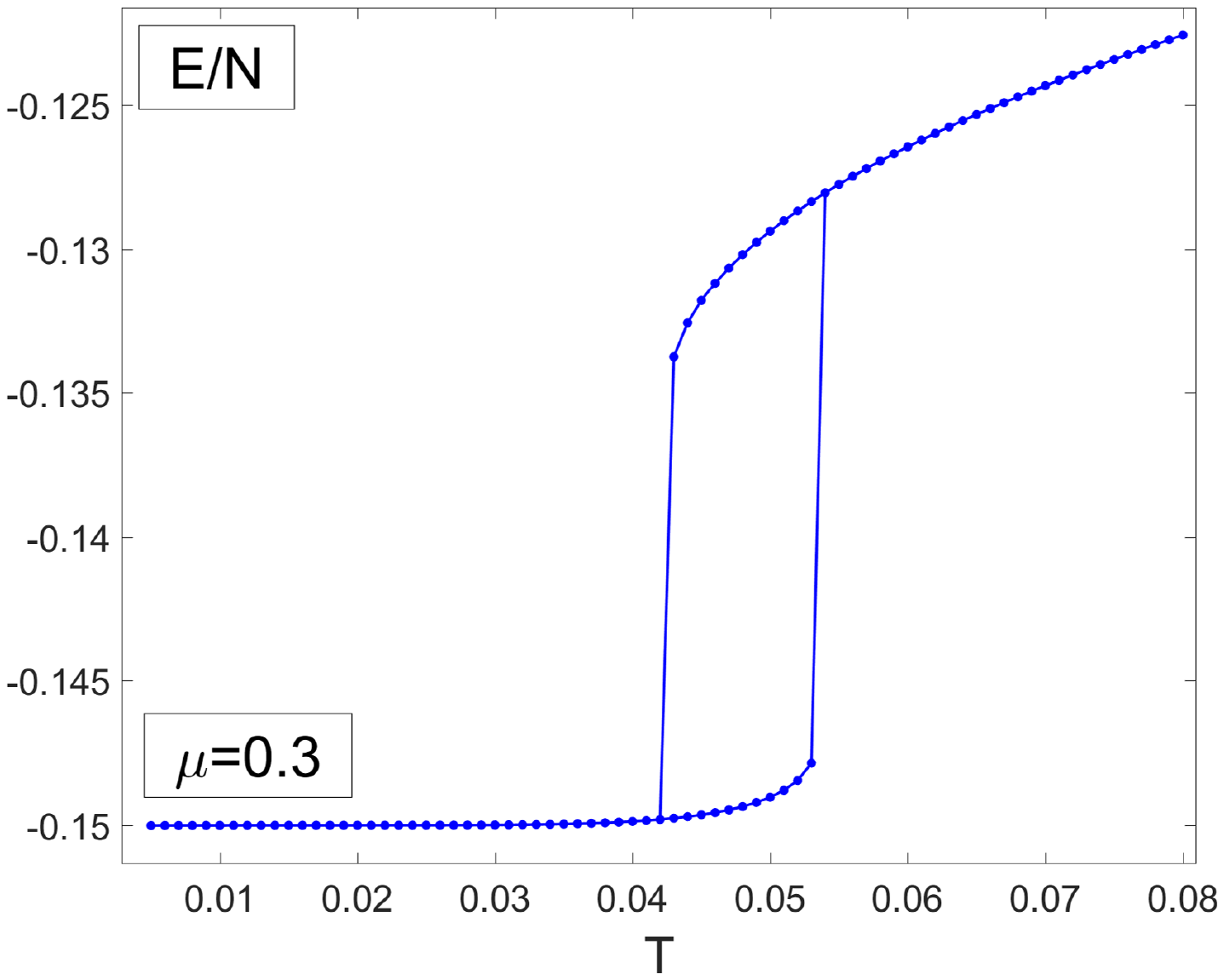}}
	\subfigure[]{\label{S-T}
		\includegraphics[height=5cm, width=7.2cm]{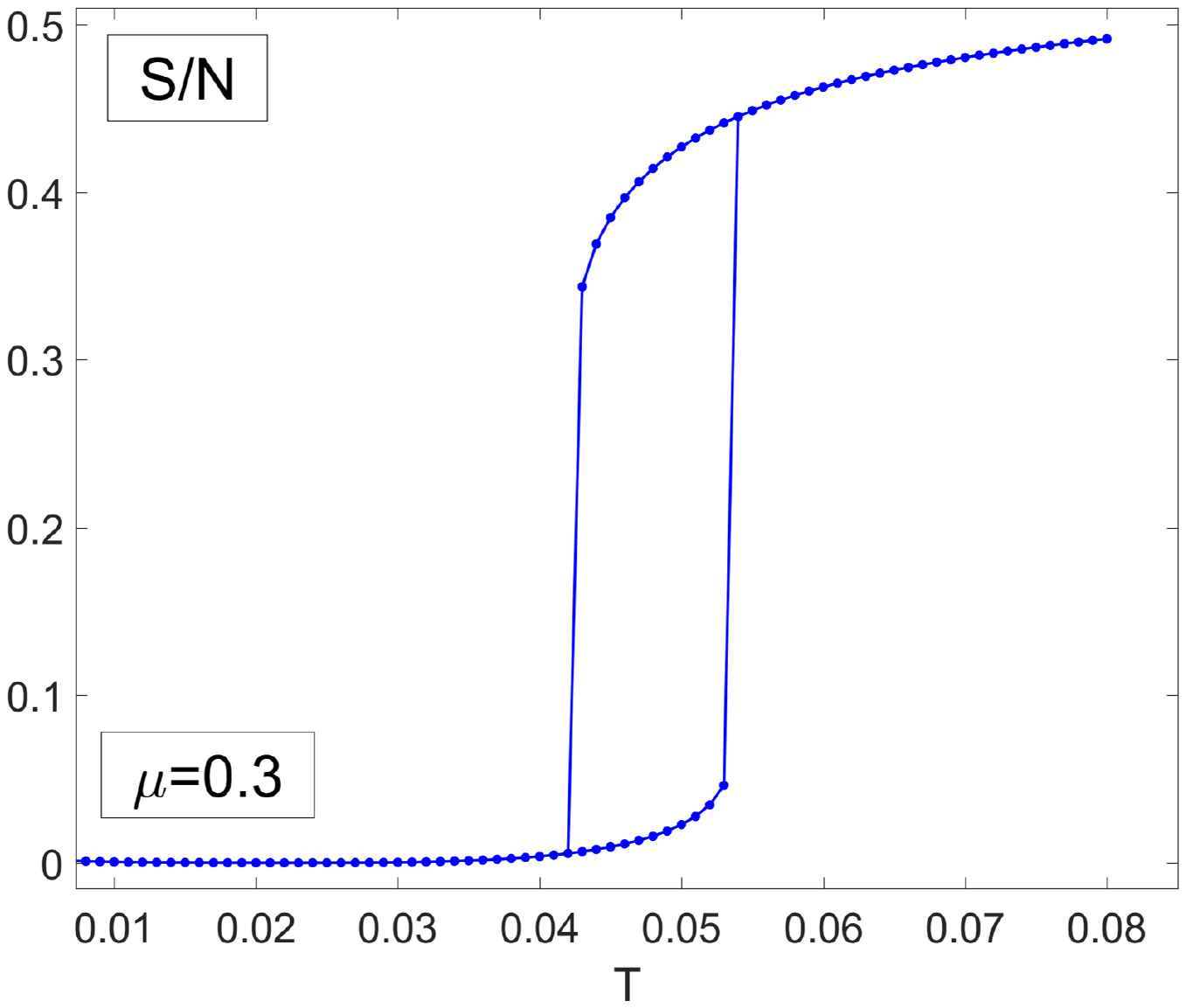}}
	\subfigure[]{\label{E-mu}
		\includegraphics[height=5cm, width=7.2cm]{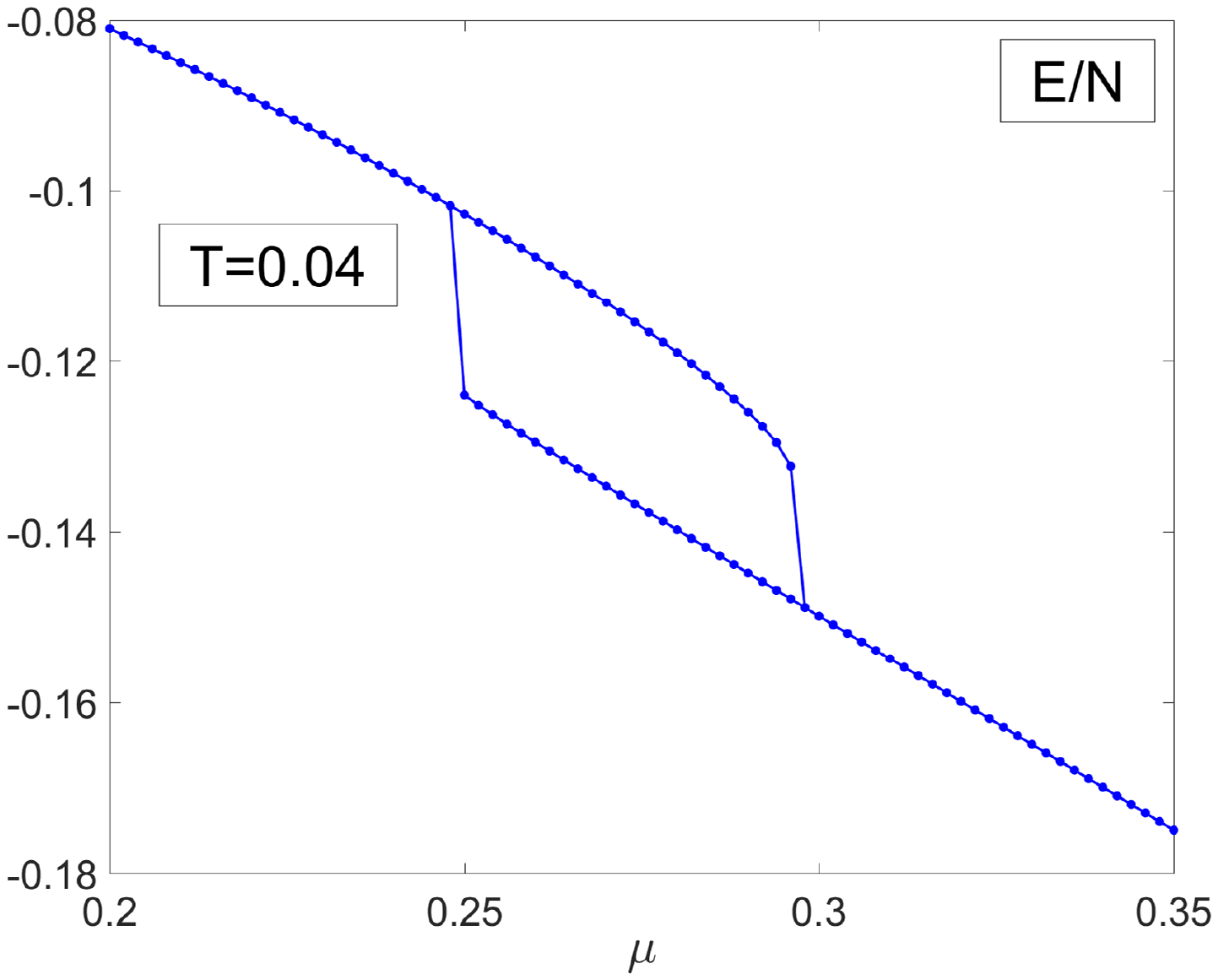}}
	\subfigure[]{\label{S-mu}
		\includegraphics[height=5cm, width=7.2cm]{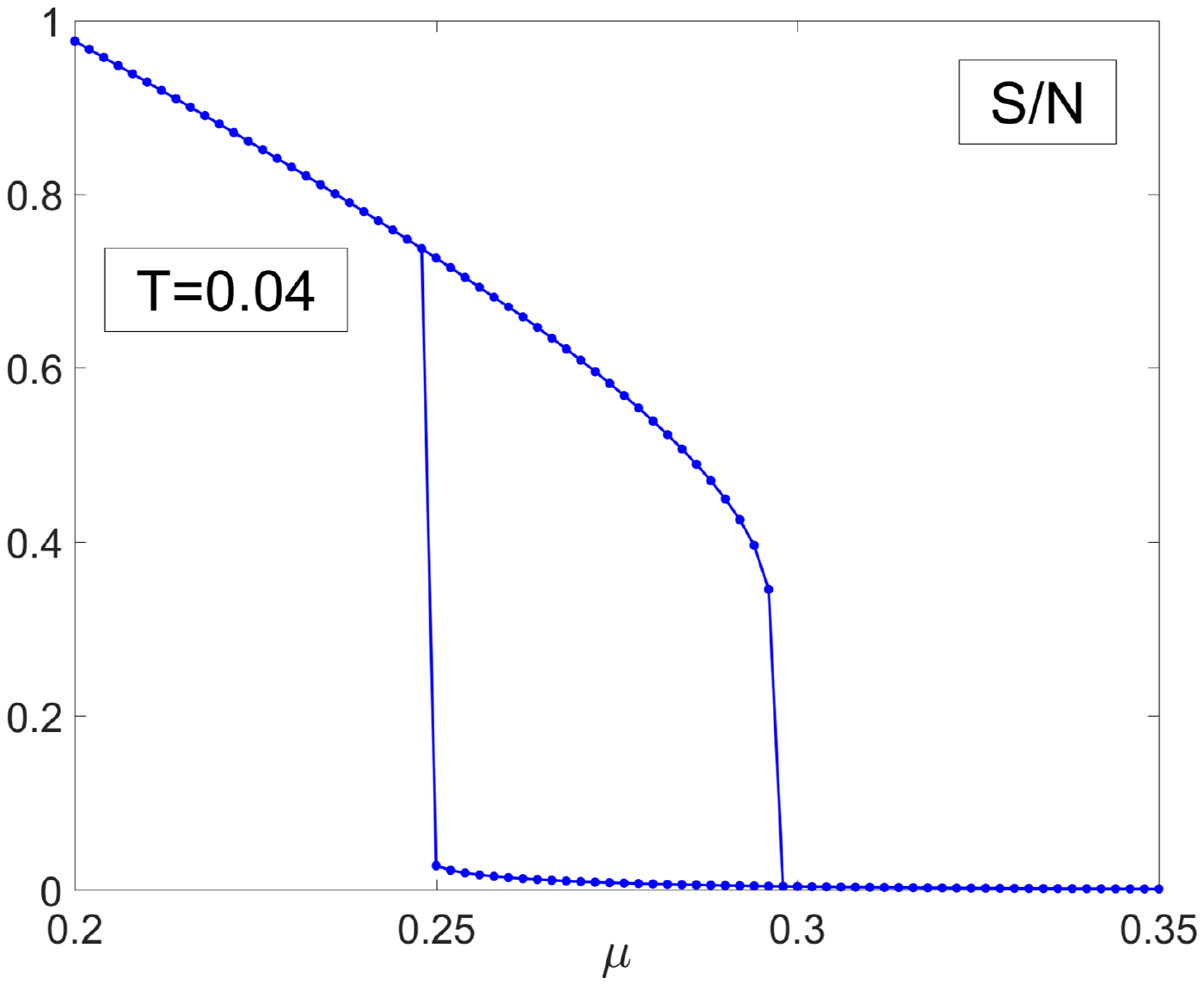}}
	\caption{(a) Thermal energy \(E/N\) as a function of temperature \(T\). (b) Entropy \(S/N\) as a function of temperature \(T\). (c) Thermal energy \(E/N\) as a function of chemical potential \(\mu\). (d) Entropy \(S/N\) as a function of chemical potential \(\mu\).}\label{energy_entropy}
\end{figure}

As depicted in Fig. (\ref{energy_entropy}a-d), we also analyze the behavior of energy and entropy as functions of temperature and chemical potential, supporting the first-order phase transition illustrated in Fig. (\ref{be1}) to (\ref{QT}). Fig. (\ref{E-T}) shows that in the low-temperature gapped phase, the thermal energy \(\frac{E}{N}\) flattens, indicating temperature independence in this phase. Both energy \(E\) and free energy \(F\) become negative as \(T \rightarrow 0\). The entropy \(S\) can be derived from \(\frac{S}{N} = \frac{1}{N} \frac{E - F}{T}\). Fig. (\ref{S-T}) demonstrates that entropy decreases to zero as temperature drops because both free energy and energy converge to the same value at low temperatures.

It's notable that the chemical potential introduces a two-body interaction. A similar interaction is found in the Maldacena-Qi wormhole model \cite{MQ18}. Comparable entropy behavior is reported in \cite{2009-10759}. Fig. (\ref{E-mu}) and (\ref{S-mu}) further investigate the impact of chemical potential on energy and entropy, clearly depicting the first-order phase transition.

\subsection{Critical Exponents}
Utilizing the numerical solutions of the Green's functions, we can explore the intricate critical behaviors of the system. Specifically, at the critical juncture, the system's temperature and chemical potential reach the values \( T_c = 0.06828 \) and \( \mu_c = 0.3443 \), respectively, marking a significant critical point.

Critical exponents, fundamental to the study of phase transitions, describe how thermodynamic quantities scale near the critical point. One of the key thermodynamic observables is the specific heat, which follows a power-law behavior close to the critical temperature. This can be expressed as
\bea
c(T, \mu_c) = T \frac{\partial S}{\partial T} \propto |T - T_c|^{-\alpha_{\pm}},
\eea
demonstrating the dramatic increase in the system's heat capacity as it approaches the critical temperature.

Charge susceptibility, another crucial observable, measures the system's response to changes in chemical potential, indicative of its ability to accumulate or deplete charge. The mathematical form derived from numerical Green's function analysis is
\bea
\chi(T, \mu_c) = \frac{\partial Q}{\partial \mu} \propto |T - T_c|^{-\gamma_{\pm}},
\eea
which signifies how the system's charge response intensifies near the critical point.

This expression encapsulates the system's sensitivity to variations in the chemical potential. Additionally, the charge density as a function of temperature is derived from a functional relationship based on our meticulous numerical analysis of Green's functions. This dependency highlights the intricate interplay between thermal excitations and electronic occupancy, illustrating how the distribution of charges within the system evolves with changing temperature. As the temperature approaches the critical value \( T_c \), the charge density profile becomes especially revealing, showcasing the characteristics of the system's phase transition and indicating potential shifts in the underlying electronic ground state. For the charge density, the critical behavior can be expressed as
\bea
|\mathcal{Q}(T, \mu_c) - \mathcal{Q}(T_c, \mu_c)| \propto |T - T_c|^{q_{\pm}}.
\eea
Moreover, the difference in charge density between two phases near the critical point behaves as
\bea\label{eq:cribeta}
\Delta \mathcal{Q} \propto |T - T_c|^{\beta_c}.
\eea
In the vicinity of the critical chemical potential \(\mu_c\), the entropy exhibits a pronounced sensitivity, providing vital insights into the mechanisms of phase transitions and the critical exponents that govern the system's thermodynamic response. The relationship for entropy is given by
\bea
|S(T_c, \mu) - S(T_c, \mu_c)| \propto |\mu - \mu_c|^{s_{\pm}},
\eea
and similarly for charge density
\bea
|\mathcal{Q}(T_c, \mu) - \mathcal{Q}(T_c, \mu_c)| \propto |\mu - \mu_c|^{\tilde{q}_{\pm}},
\eea
where \(\alpha_{+}\) denotes \(\alpha\) computed above the critical point (\(T > T_c\)) and \(\alpha_{-}\) denotes \(\alpha\) computed below the critical point (\(T < T_c\)), with similar notation for \(\gamma_{\pm}\), \(q_{\pm}\), \(s_{\pm}\), and \(\tilde{q}_{\pm}\).
In close proximity to the critical coordinates \( T_c \) and \(\mu_c\), leveraging the numerical techniques expounded upon in the seminal work of Ferrari et al. \cite{Ferrari}, we employ advanced computational methods to accurately estimate the critical exponents of the model. These methodologies have been instrumental in generating the data summarized in Table \ref{table}, which showcases the essential scaling properties of the system at its critical state.

\begin{table}[!h]
	\centering
	\caption{Critical exponents}\label{table}
	\begin{tabular}{c|l|c|l}
		\hline
		\hline
		$~~~\alpha_{+}~~~$& ~~~0.6388~~~ &  	$~~~\alpha_{-}~~~$&  ~~~0.6644~~~\\ \hline
		$\gamma_{+}$& ~~~0.5815~~~ & 	$\gamma_{-}$ & ~~~0.7537~~~ \\ \hline
		$q_{+}$& ~~~0.4007~~~ &$q_{-}$  & ~~~0.4714~~~ \\ \hline
		$s_{+}$& ~~~0.5202~~~ &$s_{-}$  & ~~~0.4224~~~ \\ \hline
		$\tilde{q}_{+}$& ~~~0.5116~~~ & $\tilde{q}_{-}$ & ~~~0.4265~~~ \\ \hline
		$\beta_c$&~~~0.6397~~~ & &  \\ \hline
	\end{tabular}
\end{table}

In the study of critical behavior in the SYK model, solving the SD equation to determine the critical exponents is a crucial step. Initially, a first-order phase transition is induced by setting the chemical potential to \(\mu = 0.34\) with an initial factor of unity (i.e., 1). The free energy \(F(T)\) is then computed using the Green's function formalism, starting from the highest temperature \(T = 0.1\) with an initial Green's function value of \(G(\tau) = 0.5\). As the temperature decreases, a small temperature step \(T_{\text{step}} = 10^{-5}\) is applied, using the Green's function from the previous iteration as the starting point for the new calculation. The free energy \(F(T)\) is evaluated at each step until the temperature reaches \(T = 0.005\), yielding the temperature-dependent free energy function \(F_-(T)\), with the chemical potential held fixed. Subsequently, the temperature is incremented in reverse from \(T = 0.005\) to \(T = 0.1\), producing the complementary free energy function \(F_+(T)\), which serves as the counterpart to the previously obtained function.

To ensure numerical convergence, a convergence check is introduced: the difference between \(F_+(T)\) and \(F_-(T)\) is calculated. If the maximum difference, \(\max(|F_+(T) - F_-(T)|)\), is smaller than \(10^{-9}\), the chemical potential \(\mu\) is updated by adding a scaled version of the maximum difference and multiplying \(\mu\) by 1.1. If the maximum difference \(\max(|F_+(T) - F_-(T)|)\) is less than or equal to \(10^{-9}\), the chemical potential is reduced by a factor of 0.5. This process is repeated until the condition
\[
10^{-9} < \max(|F_+(T) - F_-(T)|) < 1.5 \times 10^{-9}
\]
is satisfied, at which point the critical chemical potential \(\mu_c = \mu\) is determined. Once the critical chemical potential is established, the next step is to identify the critical temperature \(T_c\), which is determined as the temperature at which the absolute value of the second derivative of \(F_+(T)\) is maximized. Next, by gradually lowering the temperature and calculating \(\Delta \mathcal{Q}\), the power-law relationship for the critical exponents can be uncovered. Specifically, at each fixed temperature, the two distinct branches of the \(F(T) - \mu\) and \(\mathcal{Q}(T) - \mu\) relations are solved, and the chemical potential is varied in opposite directions. The point where the two branches intersect corresponds to the chemical potential \(\mu_i\) at which a first-order phase transition occurs. The quantity \(\Delta \mathcal{Q}(T)\) is then computed by taking the absolute difference between the \(\mathcal{Q}\) values at \(\mu_i\) for the two branches:
\[
\Delta \mathcal{Q}(T) = \left|\mathcal{Q}_{\text{increasing} \mu}(\mu_i) - \mathcal{Q}_{\text{decreasing} \mu}(\mu_i)\right|.
\]
By varying the temperature step by step and repeating this procedure, the power-law relationship
\(
\Delta \mathcal{Q} \propto |T - T_c|^{q_\pm}
\)
can be revealed, which provides a detailed quantitative analysis of the phase transitions and critical phenomena in the SYK model. Through this series of precise calculations and analyses, the critical exponents can be accurately determined, offering profound insight into the phase transition behaviors and critical phenomena in the SYK model.


\begin{figure}[t]
	{
		\includegraphics[height=4.4cm, width=5.6cm]{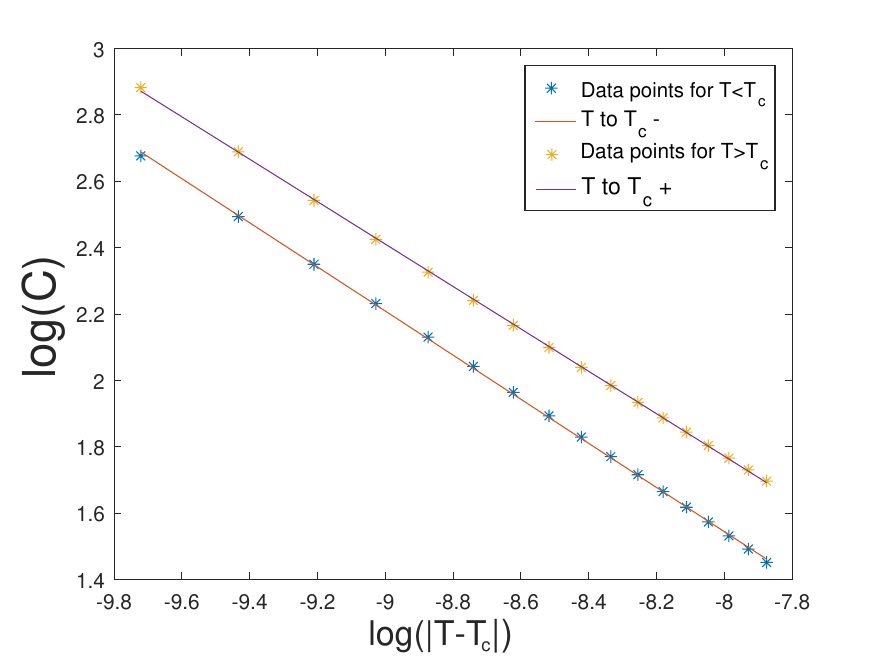}}\label{fig3}\quad\quad
{\includegraphics[height=4.4cm, width=5.6cm]{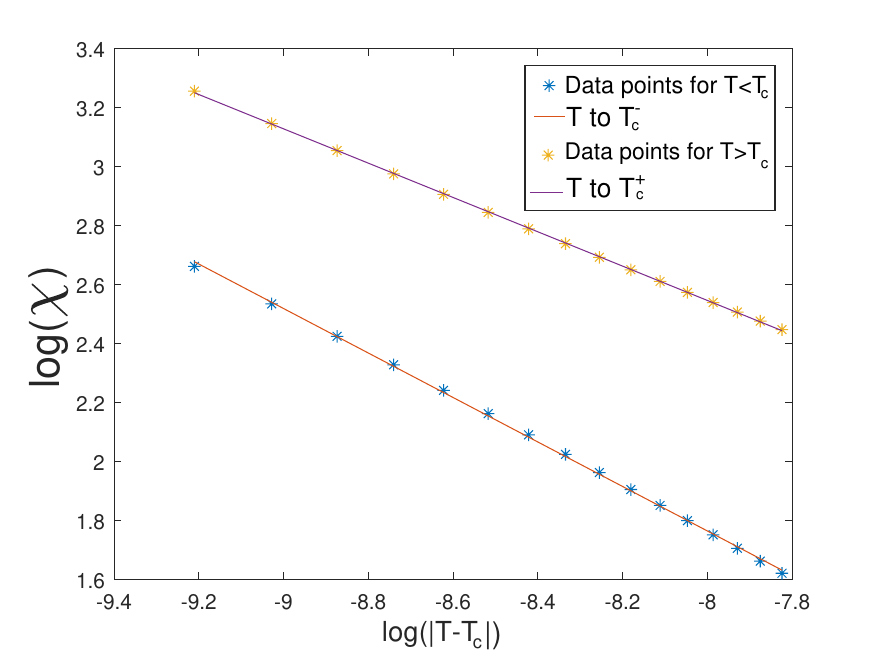}}\quad\quad
{
\includegraphics[height=4.4cm, width=5.6cm]{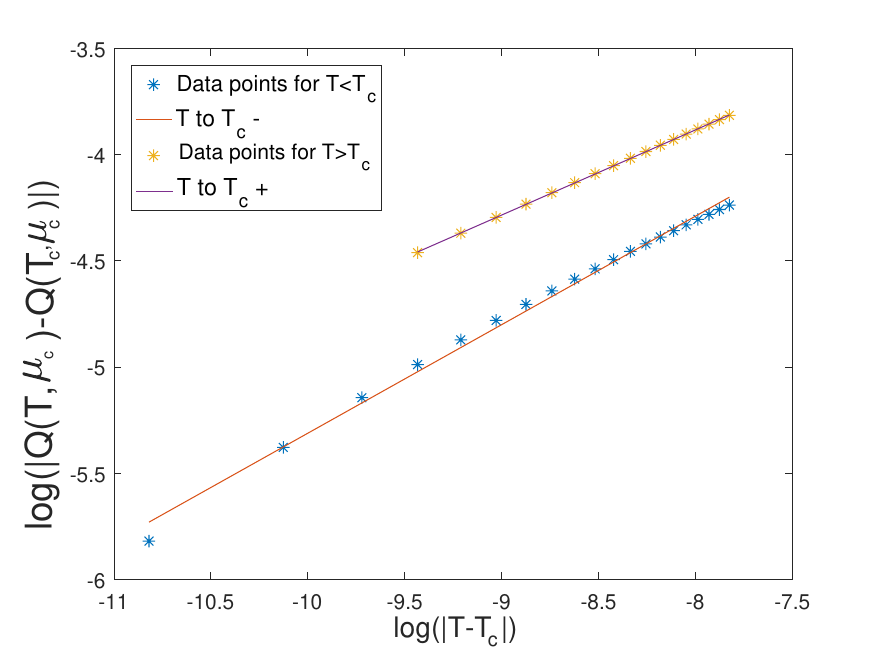}}\quad\quad
{
\includegraphics[height=4.4cm, width=5.6cm]{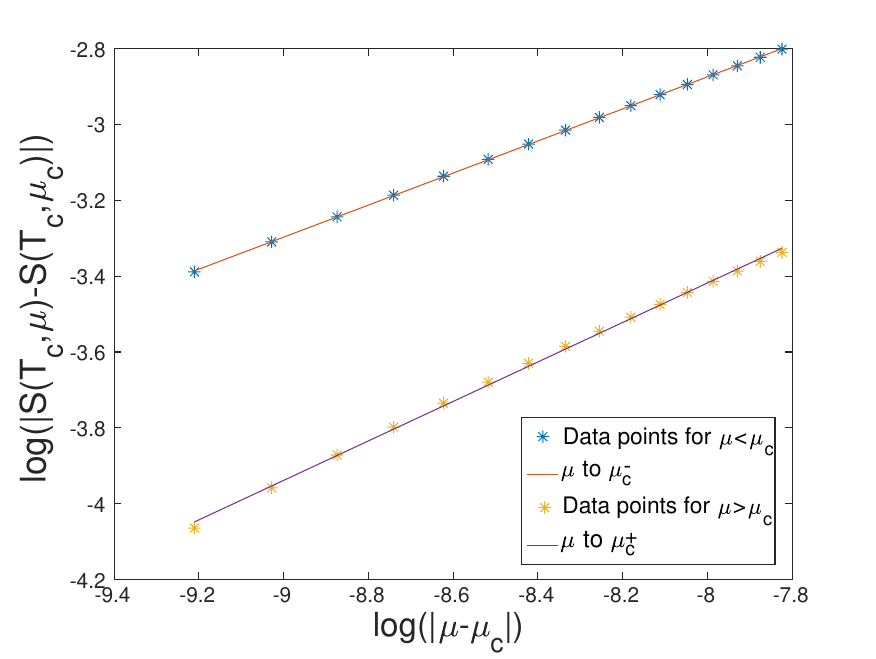}}\quad\quad
{
\includegraphics[height=4.4cm, width=5.6cm]{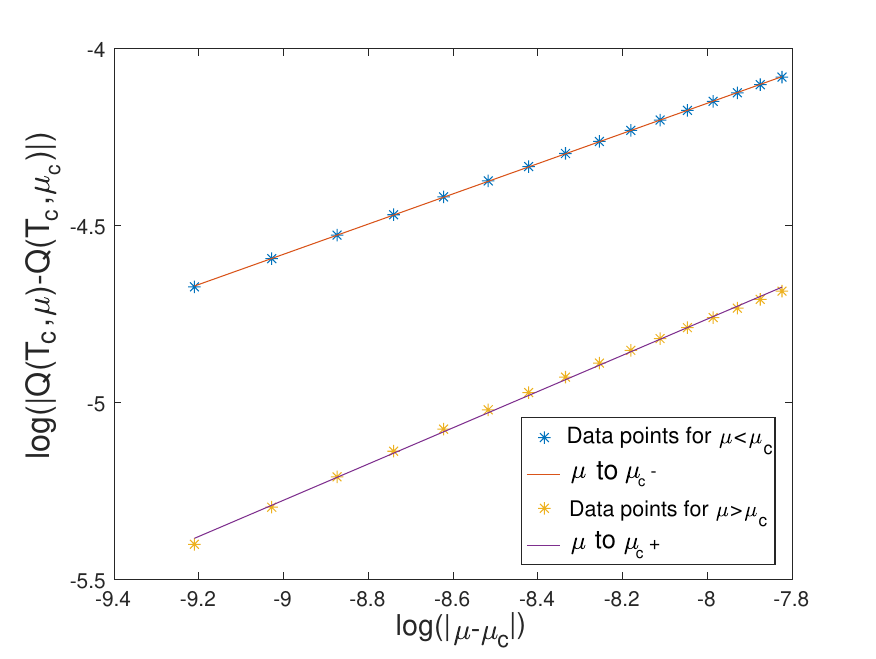}}\quad\quad
{
\includegraphics[height=4.4cm, width=5.6cm]{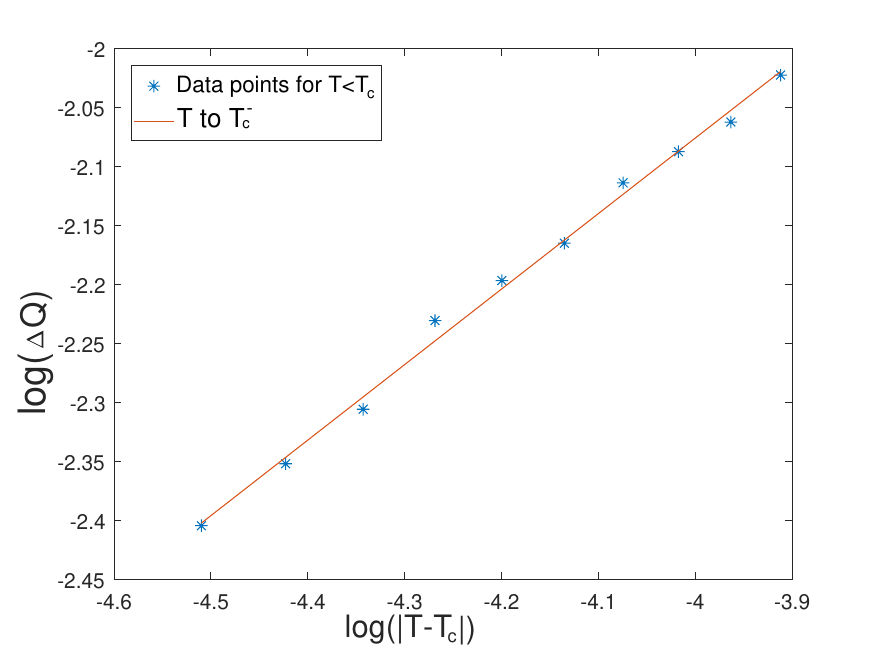}}\quad\quad
\caption{Behaviors of thermodynamic functions near the critical point.}\label{CE1}
\end{figure}

The critical exponents derived from our study highlight that the complex cSYK model can act as a microscopic statistical representation of the classical liquid-gas phase transition. Notably, Ferrari et al. provided a comprehensive examination of non-mean field criticality in cSYK models in \cite{Ferrari}. Their analysis not only addresses the original cSYK model but also explores two alternative variants, focusing on their thermodynamic properties. Interestingly, they demonstrate that this non-mean field behavior persists even in certain higher-q instances beyond the \(q=4\) case. The core insight from their research is that the effective action of the cSYK model depends on functional relationships rather than a finite set of variables, which leads to these unconventional critical phenomena. This discussion specifically pertains to the four-body interaction scenario, or the \(q=4\) variant of the cSYK model. Contrastingly, another significant finding reported in \cite{2204-09629} shows that a suitably parameterized many-body SYK model can exhibit mean-field theory behavior and present a Van der Waals-like phase diagram when thermodynamic quantities are appropriately scaled. This broadens the applicability of SYK models in describing diverse phase transitions, bridging the gap between non-mean field and mean-field behaviors in quantum systems.

We present our numerical calculations on the critical exponents \(\alpha_{\pm},~\gamma_{\pm},~q_{\pm},~\beta_{\pm},~s_{\pm}\), and \(\tilde{q}_{\pm}\) as shown in Fig. (\ref{CE1}). It is worth noting that \(\beta_c\) obtained in our calculations shows a precision difference of up to \(10^{-9}\) compared to the results from \cite{Ferrari}.

This section's thermodynamic analysis has revealed the characteristics of phase transitions driven by the chemical potential in the complex Sachdev-Ye-Kitaev model. We have discovered that variations in the chemical potential not only affect the system's phase state but also significantly regulate the distribution of energy levels and the system's entropy behavior. These thermodynamic properties provide a crucial theoretical foundation for further exploration of the system's quantum chaos features and energy level dynamics. Due to the varying influence of the chemical potential, the system can be divided into two distinctly different phases. Taking this as an inspiration, we will next employ the exact diagonalization method to precisely study the energy level distribution and chaotic behavior of the cSYK model and further explore how the chemical potential, which induces thermodynamic phase transitions, specifically affects their behavior.

\section{Exact diagonalization and spectral form factor}\label{ED}

\begin{figure}[t]
    \centering
    \subfigure[]{\includegraphics[width=0.319\textwidth]{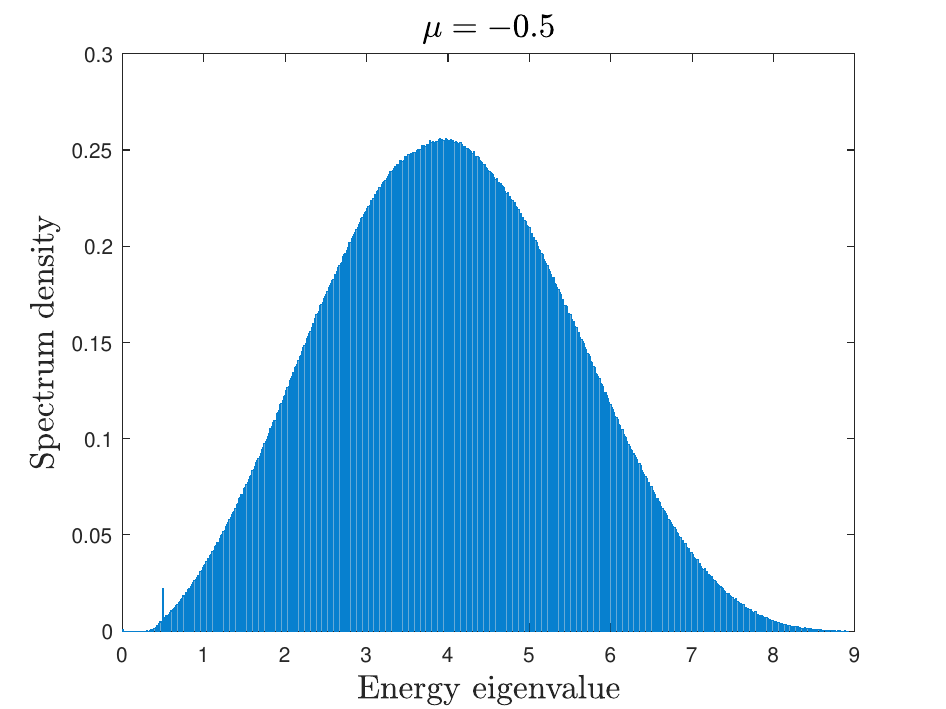}}
    \hspace{0cm} 
    \subfigure[]{\includegraphics[width=0.319\textwidth]{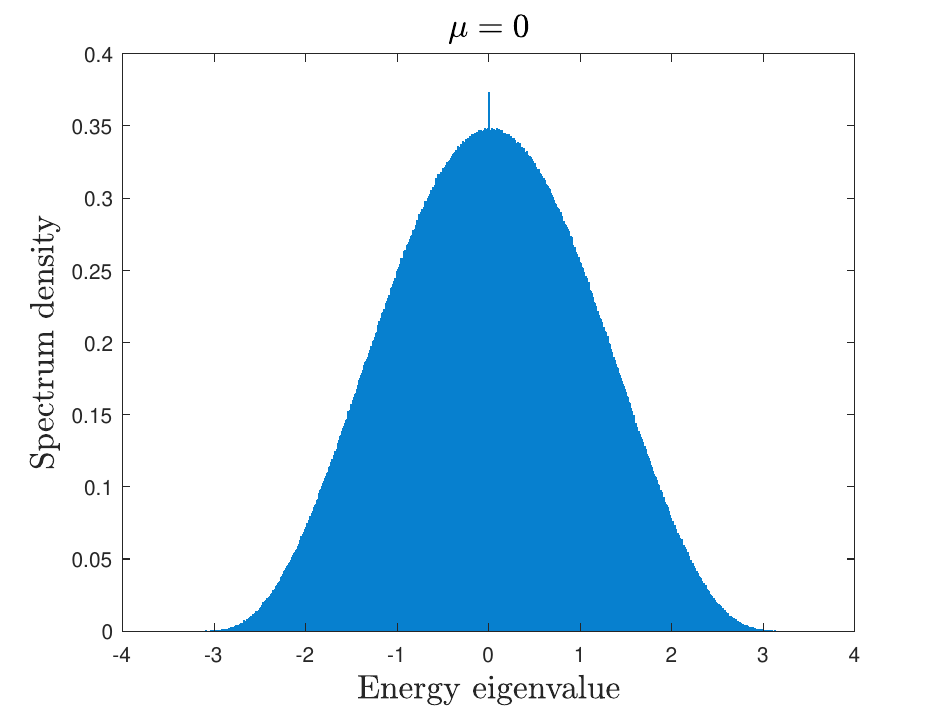}}
    \subfigure[]{\includegraphics[width=0.319\textwidth]{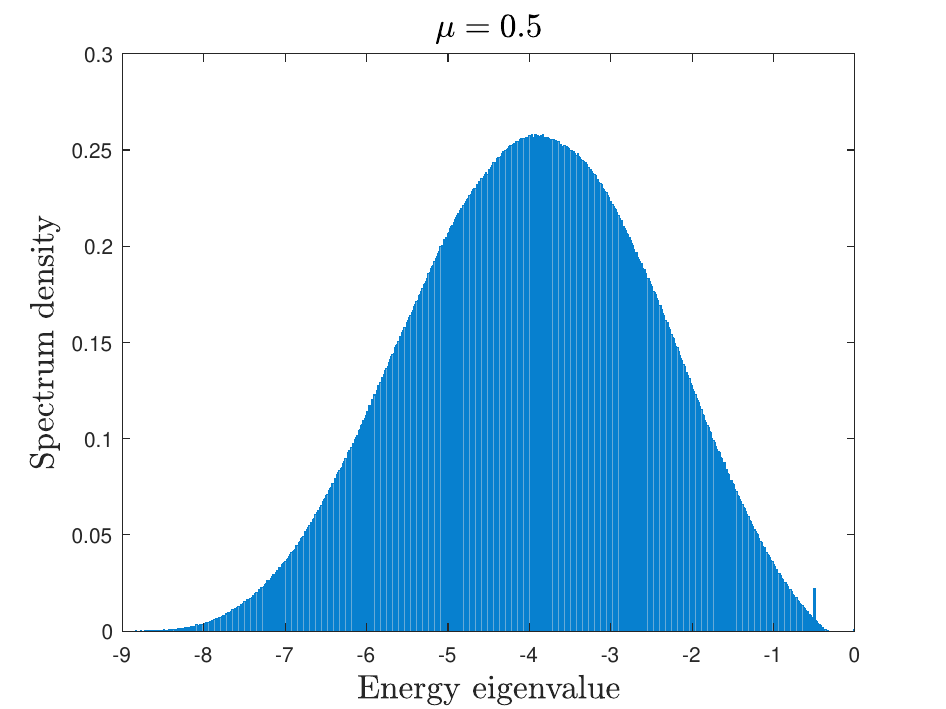}}
    \caption{Spectrum Density with the average of 20 samples at $N=16$. The chemical potential obviously breaks the symmetry of the energy spectrum  renders a skewed distribution.}
\label{spectrum}
\end{figure}

\begin{figure}[t]
    \centering
    \subfigure[]{\includegraphics[width=0.49\textwidth]{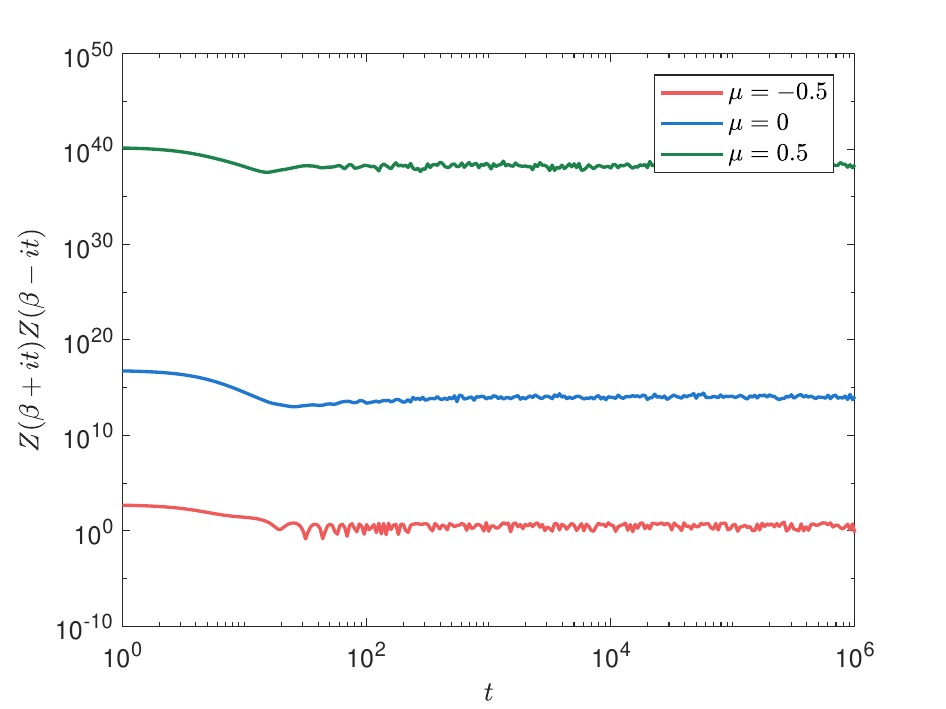}}
    \hspace{0cm} 
    \subfigure[]{\includegraphics[width=0.49\textwidth]{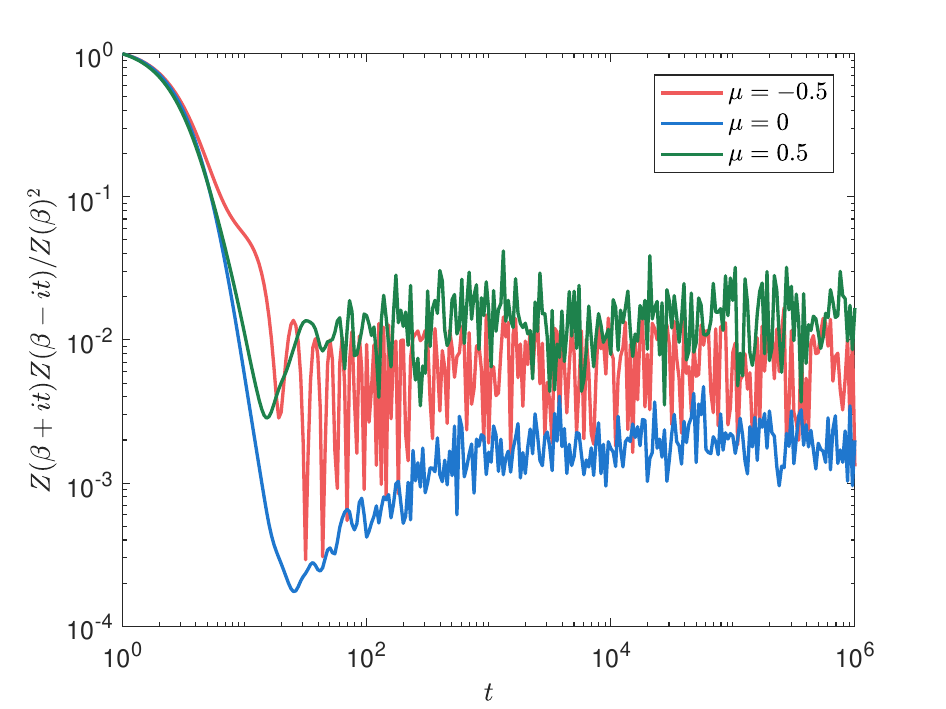}}
    \caption{(a) The spectral form factor with chemical potential $\mu=-0.5,~0,~0.5$, the chemical potential significantly affects the order of magnitude of SFF. (b) Regularized spectral form factor with the average of 20 samples at $N=16$, where $\beta=5$. The fluctuation of plateau stage of neutral case is obviously less than the others.}\label{SFF}
\end{figure}
In addition to the study of thermodynamic properties, it is essential to delve deeper into the spectral characteristics and dynamical behavior of the cSYK model. While the previous sections have focused on understanding the thermodynamic phase transitions and critical phenomena, exploring the spectrum and real-time dynamics can provide further insights into the microscopic details of the model. In this section, we present the numerical results obtained from the exact diagonalization method for the cSYK model with $N=16$ \cite{Fu_2016,Cotler_2017}. The focus will be on the spectrum density and the SFF for different values of the chemical potential $\mu$. These investigations are crucial for a comprehensive understanding of the system's behavior. To achieve this, we employ exact diagonalization method to analyze the spectrum density and spectral form factor (SFF) of the cSYK model. These tools allow us to examine the eigenvalue distribution. By investigating the concrete finite size model, we can gain a clear and deep understanding of the quantum chaos and the statistical properties of the eigenvalues in different phases of the model. The spectral form factor, in particular, serves as a valuable indicator of quantum chaotic behavior and can reveal important information about the dynamics and thermalization processes in the system. Specifically, we employee the Jordan-Wigner transformation to express the fermion operator as a series of Pauli string
\begin{align}
    c_i=\sigma_i^-\bigotimes_{j>i}\sigma^z_j~,\quad
    c_i^\dagger=\sigma_i^+\bigotimes_{j>i}\sigma^z_j~, 
\end{align}
notice there should be a identity operator $I_2^{\otimes^{i-1}}$ in front of the Pauli operator $\sigma^{\pm}$ to guarantee the operation is over the entire Hilbert space. Through these concrete Pauli matrices, the physical behavior of interest can be studied by doing various operations on it, such as diagonalizing the Hamilton matrix to obtain the energy spectrum or numerically calculating the exponential of Hamiltonian matrix to derive the time evolution operator $U(t)=e^{-iHt}$ and so on. 
The spectrum density provides valuable information about the distribution of energy eigenvalues, which is crucial for understanding the statistical properties of the system. Fig. (\ref{spectrum}) shows the spectrum density for $\mu = -0.5$, $\mu = 0$, and $\mu = 0.5$, averaged over 100 samples.
\begin{itemize}
    \item For $\mu = -0.5$, the spectrum density exhibits a maximum around $E \approx 4$, indicating a higher concentration of energy eigenvalues in this region. The distribution shows a slight skewness towards higher energy values.
    \item For $\mu = 0$, the spectrum density is symmetric and maximum around $E = 0$. This symmetry and the central peak suggest that the system is neutral, with an equal distribution of positive and negative energy states.
    \item For $\mu = 0.5$, the spectrum density has a maximum around $E \approx -4$. Similar to the $\mu = -0.5$ case, but in the opposite direction, the distribution is skewed towards lower energy values.
\end{itemize}
These behaviours indicate that the chemical potential $\mu$ significantly affects the energy distribution, shifting the peak of the spectrum density accordingly. For all three case of chemical potential. Furthermore, the spectrum density can always concentrate a peak at $E=\mu$ due to the presence of the conserved charge $\mathcal{Q}$ which is not observed in regular Majorana SYK model. The peak at $E = \mu$ suggests that the system has a significant number of states available at this energy, which can be attributed to the alignment of energy levels with the chemical potential. This behavior highlights the role of the chemical potential in shaping the energy distribution and hints the potential localization or phase transition effects in the system. Another subtle detail is that there exists a energy gap of $\mu=\pm0.5$ cases virtually. All the samples of $\mu=\pm0.5$ possess the energy level $E=0$, and the energy level space between $E=0$ and the next($\mu=-0.5$) or previous($\mu=0.5$) energy level are significantly larger than the other energy levels.

Next, we analyze the spectrum form factor, which is a powerful tool for studying quantum chaos and the dynamical properties of the system. The SFF is defined as
\begin{align}
    F(t) = Z(\beta + it)Z(\beta - it)=\sum_{m,n}e^{-\beta(E_m+E_n)+it(E_m-E_n)},
\end{align}
where $Z(\beta + it)$ is the partition function at complex time. In Fig. (\ref{SFF}a), the amplitude of unnormalized spectral form factor with chemical potential $\mu=0,
~\pm0.5$ is significantly affected by the chemical potential since the $F(t=1)$ is the square of partition function. The behaviour is generally consistent with that of the Majorana SYK model. All three cases show a rapid slope at the first stage then reach the dip at a time scale of $10^2$. After the ramp, the SFF eventually stay oscillation at the plateau stage. As shown in Fig. (\ref{SFF}b), the normalized SFF $F(t)/F(t=0)$ indicates that the system exhibits smaller fluctuations in the neutral case. This signifies that, in the absence of an external chemical potential, there are no additional particle or hole excitation, consequently resulting in relatively stable energy state.

\section{Real-time dynamics and spectral functions}\label{realtime}
\begin{figure}[t]

    \centering
    \subfigure[]{\includegraphics[width=0.49\textwidth]{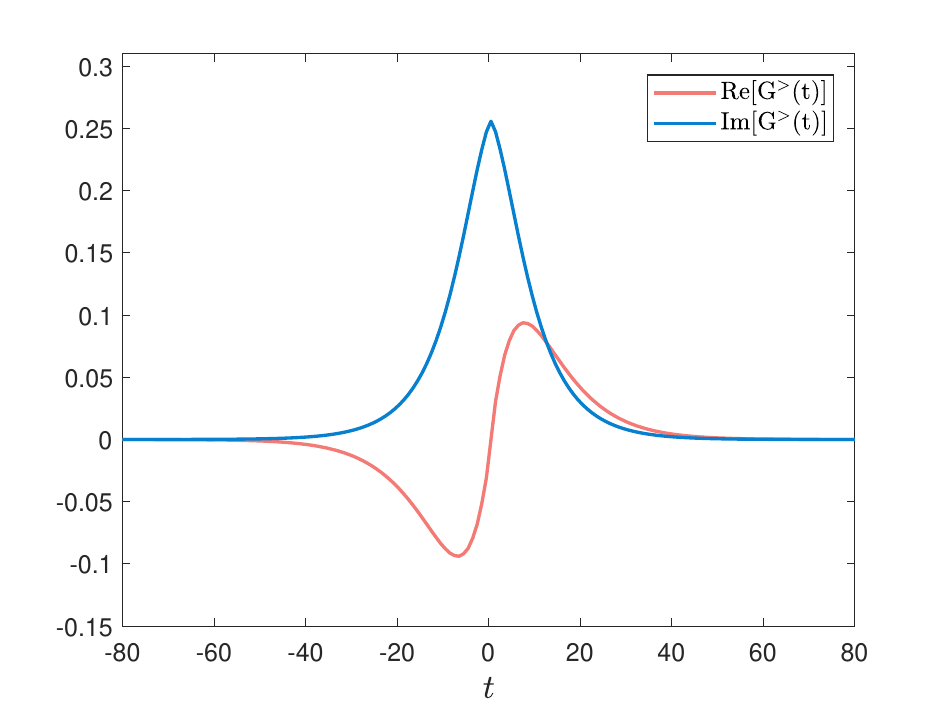}}
    \hspace{0cm} 
    \subfigure[]{\includegraphics[width=0.49\textwidth]{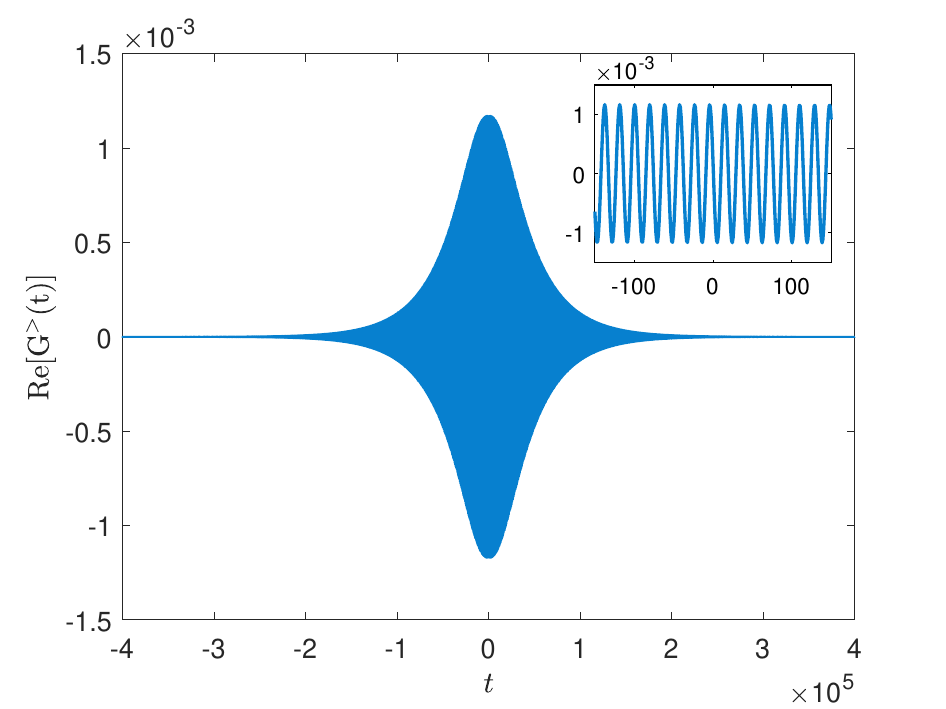}}
    
    \caption{(a) The real part and imaginary part of the gapless phase greater Green's function where $T=0.055,~\mu=0$. (b) The real part of gapped phase Greater Green's function where $T=0.055,~\mu=0.33$, the subgraph in the upper right represents a local zoom of the main graph near $t=0$, a clear harmonic oscillator-like oscillation mode can be observed.}\label{real_G}
\end{figure}

\begin{figure}[t]

    \centering
    \subfigure[]{\includegraphics[width=0.49\textwidth]{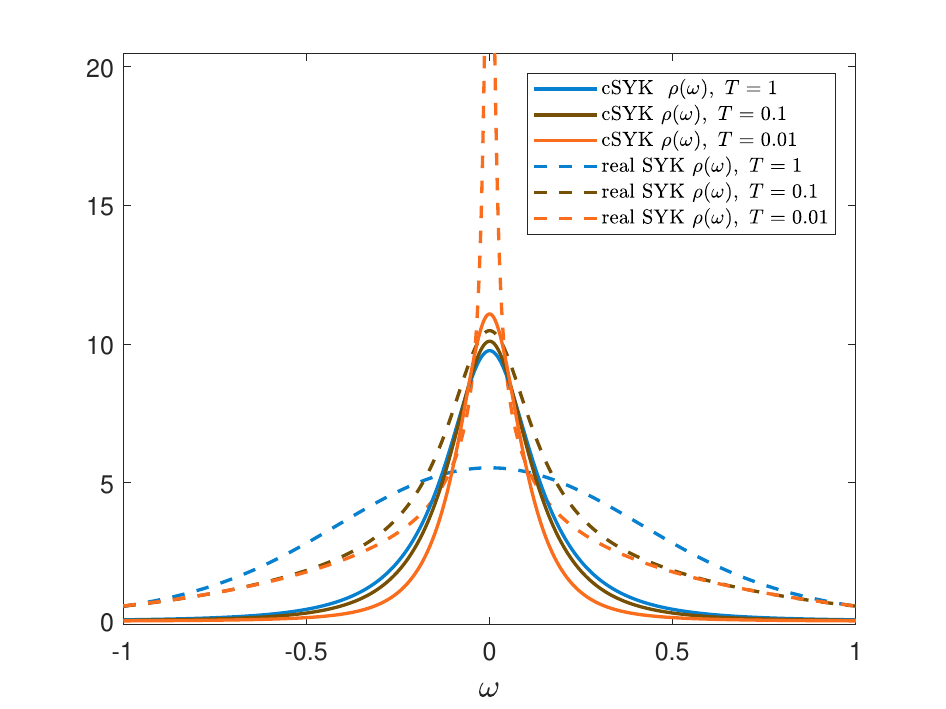}}
    \hspace{0cm} 
    \subfigure[]{\includegraphics[width=0.49\textwidth]{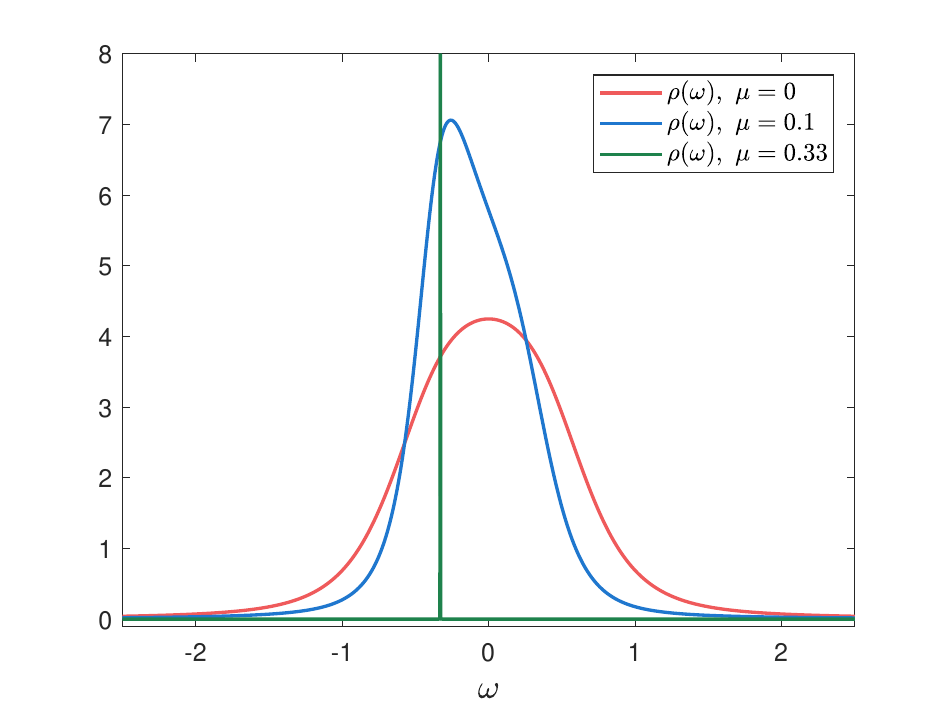}}
    
    \caption{(a) The spectral function of gapless phase of cSYK model at different temperatures (solid line) and spectral function of real SYK model as a comparison (dashed line), (b) The comparison of the spectral function between the gapless phases ($\mu=0, 0.1$) and the gapped phase of cSYK model ($\mu=0.33$), where we choose the temperature $T=0.055$ and the spacing of frequency is as low as $10^{-6}$ to ensure the numerical accuracy of the gapped phase.}\label{rho}
\end{figure}
In the real time dynamics, the behavior of the model also has two completely different sets of patterns \cite{2003-03914,2009-10759,Jian2021,Maldacena2021}. Through the analytically continuation of Euclidean time and frequency $\tau\rightarrow i t$, $\omega_n\rightarrow i\omega$, we derive the real time and real frequency SD equations
\begin{equation}
 \label{real_SD}
G^{R}(\omega)=\frac{1}{-\omega-\mu-\Sigma^{R}(\omega)},\quad
\Sigma^>(t)=-J^2 G^>(t)^2G^>(-t),
\end{equation}
where the retarded Green functions and self energy are defined by
\begin{align}
    &G^R(t)=\theta(t)(G^>(t)-G^<(t)),\\
    &\Sigma^R(t)=\theta(t)(\Sigma^>(t)+\Sigma^<(t)).
\end{align}
Further, the spectral function can be extracted from the retarded Green's function
\begin{equation}\label{rho_func}
    \rho(\omega)=-2{\rm Im}\left[
    G^R(\omega)
    \right].
\end{equation}
The iteration algorithm we explained before is also available for above functions. But the conditions is slightly different here, different with the Fourier series transformation under the Euclidean signature, in Lorenzian signature, the integral of the Fourier transform is continuous, with frequency and time ranging from negative infinity to infinity. 

In Fig. (\ref{real_G}), we compare the behaviour of greater Green's function gapless and gapped phase. Fig. (\ref{real_G}a) shows a rapid disappearance of propagator in the absence of chemical potential after the excitation. On the other hand, for the gapped phase where the chemical potential is not vanishing, the oscillation lifetime of propagators is significantly enhanced from the original $10^1$ to $10^5$ order of magnitude. The gapless phase dominated by chemical potential term enhances the propagation on sites. In this sense, the gapped phase is similar with a fermionic version of harmonic oscillator dynamically \cite{Qi2020,2009-10759}. We further study the behaviour of spectral functions \eqref{rho_func}. In Fig. (\ref{rho}a), the spectral function of gapless phase with no chemical potential is continuous and even-parity. The cSYK model’s spectral features with no chemical potential show a broad, Gaussian-like distribution, analogous to the spectral characteristics observed in the real SYK model under same temperature. This similarity underlines the robustness of the SYK model’s core spectral features, irrespective of the fermion type (Majorana vs. complex) when external parameters like chemical potential are neutral. So one can argue that the behavior of a cSYK model with zero formula is consistent with that of a real SYK model from this perspective, just as a zero charge RN black hole degenerates back to a Schwarzschild black hole. Overall, the distribution of the spectral function is relatively broad, but as the temperature increases, the spectrum can be narrower and higher. This indicates that the system does not have a significant energy gap between excitations. The emergence of the gapped phase signifies the presence of an energy gap in the system's low-energy region. This phenomenon results from the redistribution of the quasiparticle spectrum, caused by the combined effects of interactions and chemical potential. As shown in Fig. (\ref{rho}b), when the chemical potential increases to a certain extent, it induces a phase transition to a gapped phase, resulting in a single peak appearing in the spectral function. This phenomenon indicates a transition of the system from a continuous spectrum to discrete energy levels.

\section{Discussion and Conclusion}\label{sec:conclusion} In summary, this study has probed the intricate thermodynamic phase structures of the microscopic cSYK model.
On the cSYK front, the free energy versus temperature plot reveals a clear hysteresis pattern, emblematic of a first-order phase transition culminating at a critical point. The realm of large chemical potential  $\mu$ corresponds to an insulating, gapped phase, while the domain of small 
 $\mu$ manifests a gapless metallic phase. The distinction between these gapped and gapless regimes is underscored by the variation in charge density difference, denoted by  $\Delta \mathcal{Q}$ , and intriguingly, along the coexistence line, there exist dual possibilities for charge polarization configurations. Remarkably, this phenomenon bears striking resemblance to the archetypal liquid-gas phase transition, wherein there is no underlying symmetry change across the two phases—a feature that highlights the universal nature of the transition dynamics across various physical systems.

In addition to thermodynamic analysis, the study employed exact diagonalization for the cSYK model with $N=16$ to delve deeper into the spectral characteristics and quantum chaos indicators. The spectrum density analysis revealed that the chemical potential significantly affects the energy distribution, shifting the peak accordingly. Notably, the presence of an energy gap was observed in the gapped phase for $\mu = \pm 0.5$. The spectral form factor calculations further provided insights into quantum chaotic behavior, with the normalized SFF indicating smaller fluctuations in the neutral case, suggesting more stable energy states without additional particle or hole excitations.
Furthermore, the real-time dynamics and spectral functions were explored through analytical continuation of the SD equations. The greater Green's function showed rapid decay in the gapless phase, while in the gapped phase, the oscillation lifetime was significantly extended, akin to a fermionic harmonic oscillator. Spectral function analysis indicated a continuous and broad distribution in the gapless phase, and a single peak in the gapped phase, highlighting a transition from continuous spectrum to discrete energy levels driven by the chemical potential. These results underscore the critical role of chemical potential in shaping the system's spectral properties and energy gap formation.

Overall, this comprehensive investigation into the cSYK model's thermodynamic, spectral, and dynamical properties sheds light on the profound similarities between quantum systems and classical thermodynamic transitions, providing a deeper understanding of quantum dynamics and phase transitions in strongly correlated systems.

\section*{Acknowledgements}We would like to thank    Pengfei Zhang, Shao-Kai Jian, Jan Louw, Tian-Gang Zhou and Shao-Feng Wu for helpful discussions. We are especially grateful to Jinwu Ye for valuable  comments on the manuscript. This work is partly supported by NSFC, China (No.12275166 and No.12311540141).
\bibliographystyle{apsrev4-1}
\bibliography{refs.bib}

\appendix

\section{The effect of chemical potential in $q=2$ complex SYK model}\label{appendix1}
\begin{figure}[t]

    \centering
    \subfigure[]{\includegraphics[width=0.49\textwidth]{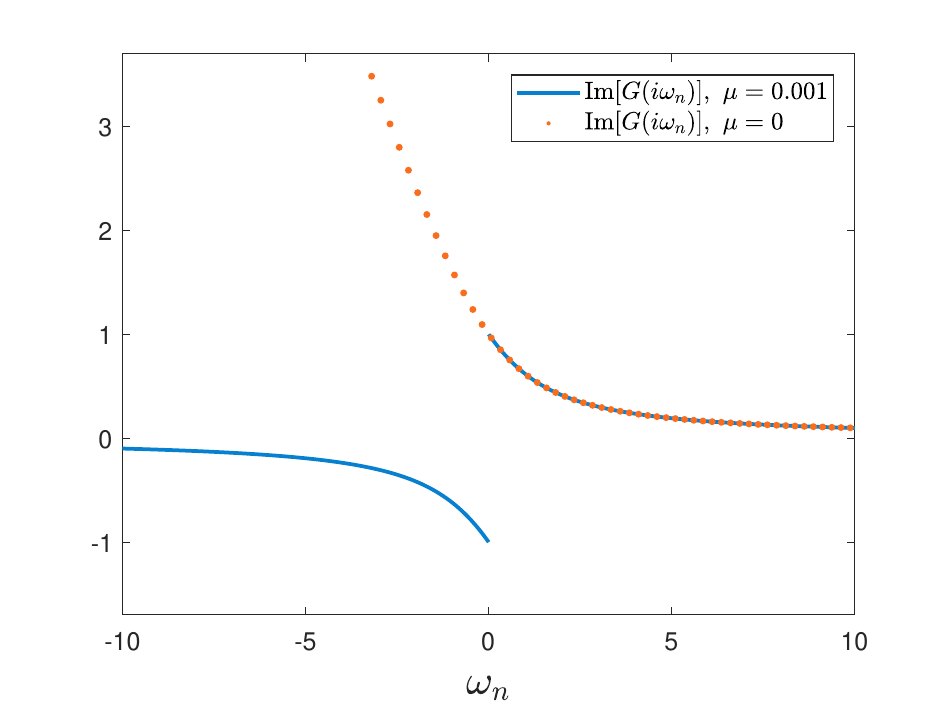}}
    \hspace{0cm} 
    \subfigure[]{\includegraphics[width=0.49\textwidth]{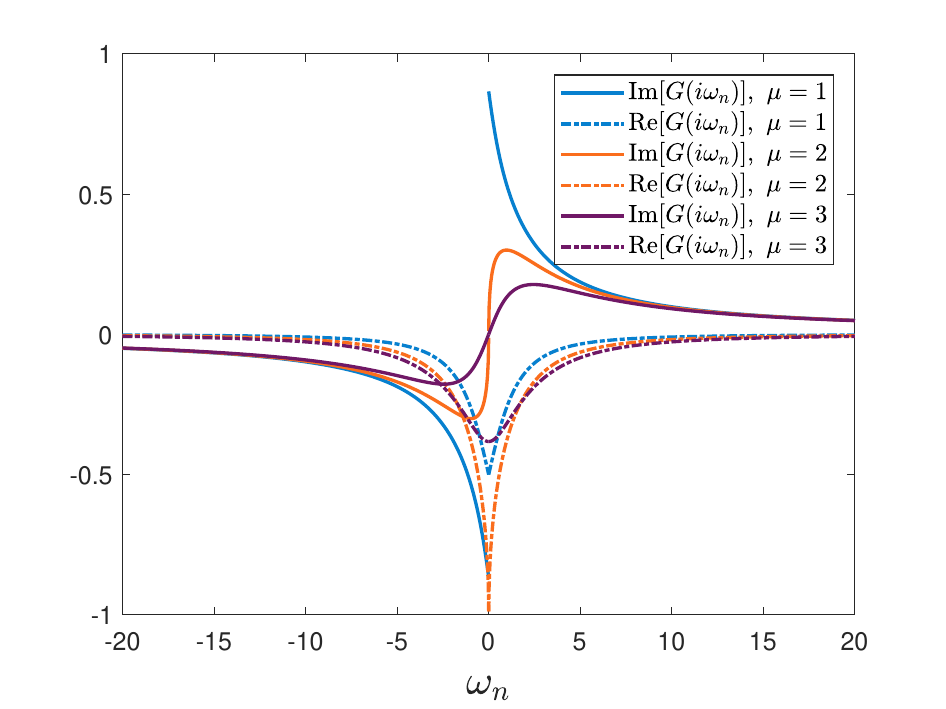}}
    
    \caption{The $q=2$ Green's function $G(i\omega_n)$ in Matsubara frequency space from the solution of \eqref{Gw_q=2} (a) with chemical potential $\mu=0,~0.001$ and (b) higher chemical potential cases. Here and below, for convenience, we set $J=1$.}\label{Gw1_q=2}
\end{figure}

In this appendix, we will focus on the analytical solution of the Schwinger-Dyson equations for the $q=2$ case to study more clearly how the chemical potential affects the behavior of the system. This will provide some implications for the phase transition behavior of unsolvable $q=4$ model. Firstly, the Schwinger-Dyson equations of $q=2$ case in Euclidean space read
\bea
\Sigma(\tau)=J^2G(\tau),\quad G(i\omega_n)=\frac{1}{-i\omega_n-\mu-\Sigma(i\omega_n)}.
\eea
The analytical solutions are given by
\bea\label{Gw_q=2}
G(i\omega_n)=\frac{-i\omega_n-\mu\pm\sqrt{(i\omega_n+\mu)^2-4J^2}}{2J^2},
\eea
the solution with positive sign is valid for all cases of non-vanishing chemical potentials, whereas it is invalid for the neutral case $\mu=0$ or negative sign. As illustrated in Fig. (\ref{Gw1_q=2}a) of the case of various chemical potentials, the Green's function in frequency space diverges at infinite of the neutral case $\mu=0$ which is considered as a non-physical solution. And the numerical iteration algorithm will not converge at zero chemical potential either, since we have introduced a UV cut-off of the high frequency, i.e. assuming $G(i\omega_n)=0$ for $|n|>n_{max}$. The fix is to manually introduce a sign function in front of the square root \cite{stanford}
\bea
G(i\omega_n,\mu=0)=\frac{-i\omega_n+{\rm sgn}(\omega)\sqrt{-\omega_n^2-4J^2}}{2J^2}.
\eea

But more importantly, no matter how small the chemical potential, the original diverging non-physical solution can naturally converge. This implies that in the cSYK model, by adjusting the chemical potential, it is possible to transform some originally non-physical solutions into physical ones. 

\begin{figure}[t]
    \centering
    \subfigure[]{\includegraphics[width=0.49\textwidth]{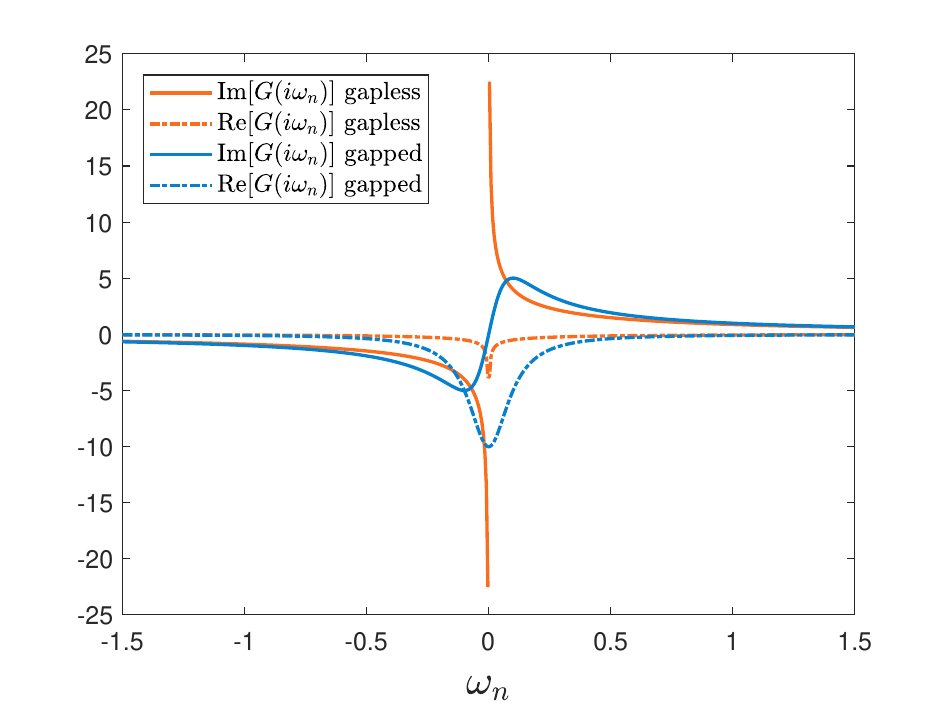}}
    \hspace{0cm} 
    \subfigure[]{\includegraphics[width=0.49\textwidth]{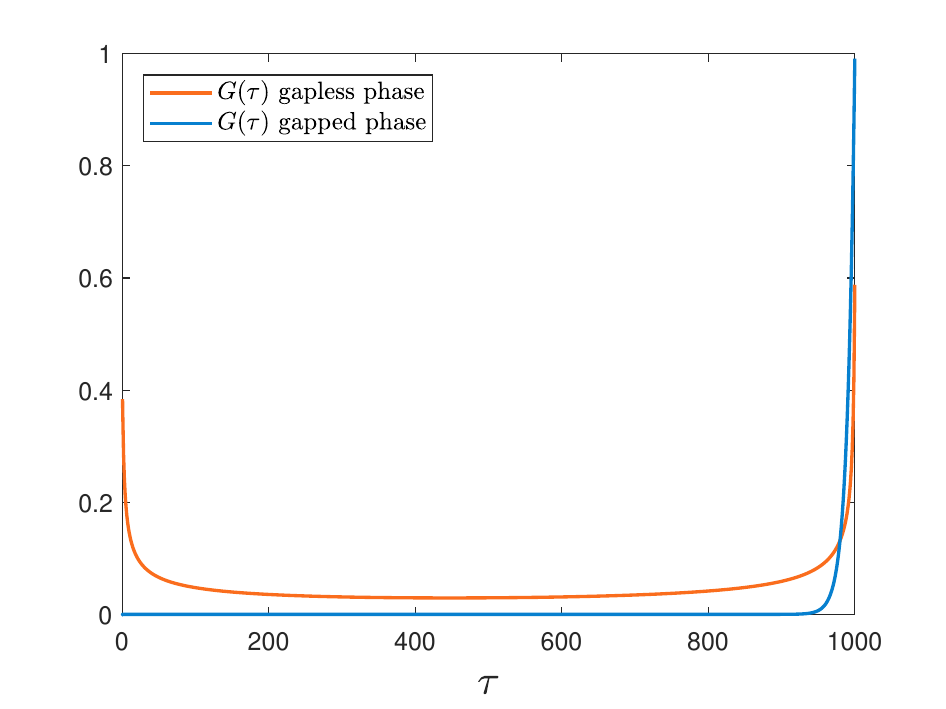}}
    \caption{(a) The numerical result of $q=4$ Green's function $G(i\omega_n)$ in Matsubara frequency space with $T=0.001, \mu=0.1$ for both gapped phase and gapless phase. (b) The corresponding gapped and gapless Green's function in imaginary time $G(\tau)$. The imaginary part of $G(\tau)$ is not shown here, because Green's function is pure real in imaginary time space for either $q=2$ or $q=4$, non-zero chemical potential or zero chemical potential.}\label{G_q=4}
\end{figure}

Additionally, one can also observe that the Green's function becomes smooth and analytic around the origin when $\mu>2J$ in Fig. (\ref{Gw1_q=2}b), and that this mechanism remains exists between the gapped and gapless phase transitions of the $q=4$ cSYK model as illustrated in Fig. (\ref{G_q=4}a). Under the same parameter $T$ and $\mu$ of gapped and gapless phase, The main characteristic difference between two different phases of Green's function is consistent with the $q=2$ cSYK model: Compared with the gapless phase, the Green's function has been smoother and analytic around the origin. For $q=2$ case, this transition is due to the analytical property nearby $\mu=2J$ of $G(i\omega_n=0^{\pm})$, i.e.
\bea
&\lim\limits_{i\omega_n\rightarrow0^+}G(i\omega_n)=\frac{-\mu+i\sqrt{4J^2-\mu^2}}{2J^2}\neq \lim\limits_{i\omega_n\rightarrow0^-}G(i\omega_n)=\frac{-\mu-i\sqrt{4J^2-\mu^2}}{2J^2}\quad {\rm for~} |\mu|<2,\nonumber\\
&\lim\limits_{i\omega_n\rightarrow0^+}G(i\omega_n)= \lim\limits_{i\omega_n\rightarrow0^-}G(i\omega_n)=\frac{-\mu+\sqrt{\mu^2-4J^2}}{2J^2}\quad {\rm for~} |\mu|\geq 2.\nonumber
\eea

\begin{figure}[t]
    \centering    \includegraphics[width=0.49\textwidth]{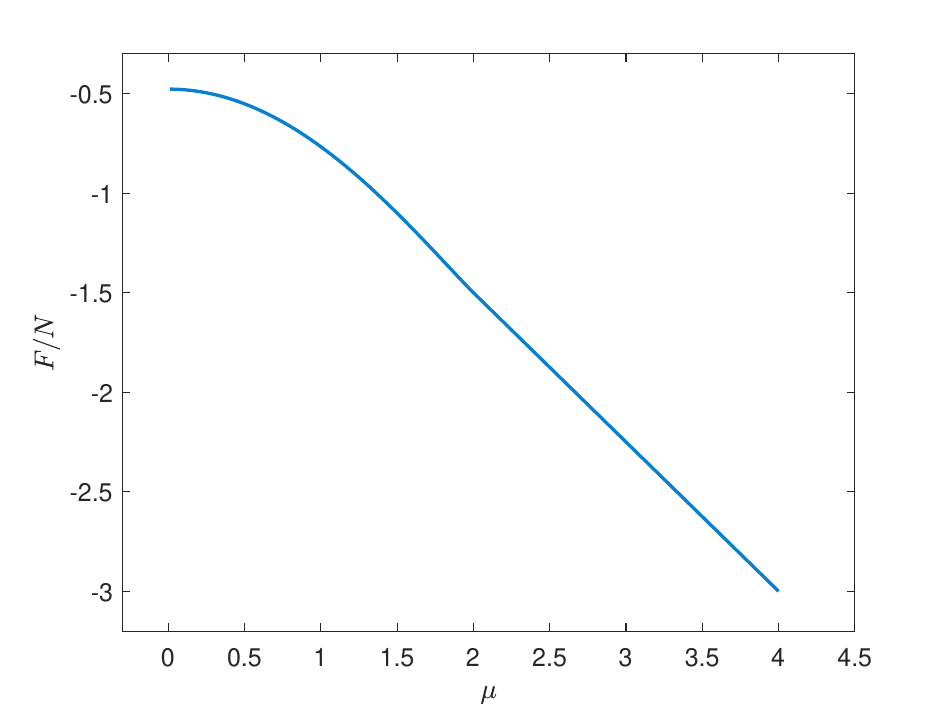}
    \hspace{0cm} 
    \caption{The free energy $F/N$ as a function a chemical potential of $q=2$ cSYK model with temperature $T=0.01$.}\label{F-mu_q=2}
\end{figure}

From this point of view, we propose that the property of the left and right limits of the Green function being equal can serve as a criterion for determining whether the system is in a gapped or gapless phase. For the case of $q=2$, when $|\mu| < 2$, the system behaves more like the real SYK model and is in the gapless phase, whereas when $|\mu| \geq 2$, it is in the gapped phase. However, since this transition is continuous and the system's SD equations do not allow multiple physically acceptable solutions for the same parameter, there should not be a sharp transition. We further investigated the behavior of the free energy for $q=2$ as a function of the chemical potential, as shown in Figure~\ref{F-mu_q=2}. As expected, the transition of free energy is smooth with no obvious turning point; as the chemical potential increases, the nonlinear behavior of the free energy gradually transitions to linear behavior, which is consistent with the numerical results presented in section \ref{sec:model}.

The mechanism behind this process lies in the chemical potential introducing an additional energy shift, which affects the system's energy level distribution and particle filling, thereby stabilizing solutions that would otherwise diverge due to a lack of energy balance. Specifically, the introduction of the chemical potential modifies the self-energy term in the Schwinger-Dyson equations, effectively suppressing the terms that cause the Green's function to diverge. In this manner, the chemical potential not only ensures the physical validity of the solutions but also provides a crucial control mechanism for the system's phase transition behavior.
\end{document}